\documentclass[review]{elsarticle}

\usepackage{lineno,hyperref}
\modulolinenumbers[5]
\usepackage{amsfonts}
\usepackage{algorithmicx}
\usepackage[ruled]{algorithm}
\usepackage{algpseudocode}
\usepackage{algpascal}
\usepackage{algc}
\usepackage{multirow}
\usepackage{amsmath}
\alglanguage{pseudocode}
\usepackage{subfig}
\usepackage{booktabs}
\journal{Computer Speech and Language}

\usepackage{multirow}
\usepackage{multicol}

\usepackage{numcompress}\biboptions{authoryear}

\usepackage{url}

\begin{document}

\begin{frontmatter}

\title{Sequential Routing Framework: \\Fully Capsule Network-based Speech Recognition}

\author[address_snu,address_sec]{Kyungmin Lee}
\author[address_snu]{Hyunwhan Joe}
\author[address_sec]{Hyeontaek Lim}
\author[address_sec]{Kwangyoun Kim}
\author[address_sec]{Sungsoo Kim}
\author[address_sec]{Chang Woo Han}
\author[address_snu]{Hong-Gee Kim\corref{corresponding}}

\cortext[corresponding]{Corresponding author}
\ead{hgkim@snu.ac.kr}
\address[address_snu]{Biomedical Knowledge Engineering Laboratory, Seoul National University, Seoul, 08826, Republic of Korea}
\address[address_sec]{Samsung Research, 56, Seongchon-gil, Seocho-gu, Seoul, 06765, Republic of Korea}

\begin{abstract}
Capsule networks (CapsNets) have recently gotten attention as a novel neural architecture.
This paper presents the sequential routing framework which we believe is the first method to adapt a CapsNet-only structure to sequence-to-sequence recognition.
Input sequences are capsulized then sliced by a window size.
Each slice is classified to a label at the corresponding time through iterative routing mechanisms.
Afterwards, losses are computed by connectionist temporal classification (CTC).
During routing, the required number of parameters can be controlled by the window size regardless of the length of sequences by sharing learnable weights across the slices.
We additionally propose a sequential dynamic routing algorithm to replace traditional dynamic routing. The proposed technique can minimize decoding speed degradation caused by the routing iterations since it can operate in a non-iterative manner without dropping accuracy.
The method achieves a 1.1\% lower word error rate at 16.9\% on the Wall Street Journal corpus compared to bidirectional long short-term memory-based CTC networks. On the TIMIT corpus, it attains a 0.7\% lower phone error rate at 17.5\% compared to convolutional neural network-based CTC networks~\citep{DBLP:conf/interspeech/ZhangPBZLBC16}.
\end{abstract}


\end{frontmatter}

\section{Introduction}
Capsule networks (CapsNets)~\citep{DBLP:conf/icann/HintonKW11, DBLP:conf/nips/SabourFH17, DBLP:conf/iclr/HintonSF18} are a kind of neural networks that represent a specific entity type with a group of neurons called a capsule instead of a single neuron.
The initial motivation of CapsNets was to abstract information explicitly by adapting an unsupervised clustering mechanism called routing-by-agreement between capsules, to conventional neural networks.
Capsules can be trained to represent not only the existence of entity types but also entity instantiation parameters such as textures, angles, colors, etc. Thus, CapsNets can be regarded as architectures for inverse graphics~\citep{DBLP:conf/nips/SabourFH17}.
CapsNets have shown higher accuracy in image classification compared to convolutional neural networks (CNNs) \citep{DBLP:conf/nips/SabourFH17, DBLP:conf/nips/HahnPK19, Malmgren1314210}.
Recently, researchers have focused more on adapting CapsNets for practical problems either by scaling their routing methods~\citep{DBLP:conf/iclr/TsaiSGS20} or by combining them with other architectures~\citep{DBLP:conf/icann/HePLHZ19, 8852016, wang-2019-towards}.
There are also attempts to apply CapsNets to classify sequence data \citep{DBLP:conf/interspeech/BaeK18, DBLP:conf/eusipco/Iqbal0KW18, DBLP:conf/icassp/WuLCLYDMHWLM19}.

Sequence to sequence (seq2seq) learning is an approach to learn mappings between sequences and is successfully implemented by neural models~\citep{DBLP:conf/nips/SutskeverVL14}.
Connectionist temporal classification (CTC)~\citep{DBLP:conf/icml/GravesFGS06} is a popular loss function in seq2seq problems.
By adapting CTC to automatic speech recognition (ASR) systems, it became possible to learn the alignments between speech signal sequences and label sequences directly unlike the conventional hidden Markov model (HMM) deep neural network (DNN) based systems~\citep{38131} which needed alignments from an HMM Gaussian mixture model (GMM) system for training the DNN.
CTC based ASR systems~\citep{DBLP:conf/icml/GravesFGS06} were first built on long short-term memory (LSTM)~\citep{DBLP:journals/neco/HochreiterS97} networks.
In order to accelerate training, CNN-based CTC networks~\citep{DBLP:conf/interspeech/ZhangPBZLBC16} were proposed. The newly proposed networks showed 2.5x increase in training speed while maintaining phoneme-level accuracies which requires relatively short-term dependencies.
Transformer-based~\citep{DBLP:conf/nips/VaswaniSPUJGKP17} ASR systems~\citep{8462506} were also proposed as a faster training model compared to recurrent seq2seq models by introducing self-attention. However, they require the whole context to compute self-attention maps for each time slice.
Recently, online ASR systems based on Transformers have been researched~\citep{9054476}.

In ASR, speech feature sequences can be regarded as two-dimensional images with the time length as width and the feature coefficient dimension as height. 
We hypothesized that CapsNets can potentially have richer representation capabilities than CNNs in ASR since the same shapes in the feature sequences represent different pronunciations depending on their positions which can be more precisely learned by CapsNets~\citep{DBLP:conf/icann/HintonKW11}.
Moreover, like CNNs, CapsNets are trained by computing the derivatives of error functions by propagating errors backwards over layers rather than input sequences, i.e., the number of steps to compute the gradients are decided according to the number of layers regardless of the length of input sequences.
Thus, given enough receptive fields, their layer-wise encoding is expected to alleviate the gradient vanishing and exploding problems that could arise when training long-term dependencies using recurrent neural networks (RNNs).

There have been attempts to apply CapsNets to seq2seq tasks in combination with other models~\citep{8852016, wang-2019-towards}, but the utilization of CapsNets could be maximized more.
In addition to the computational burden due to the routing mechanism itself~\citep{9207533}, there are hurdles in adopting existing CapsNets to seq2seq problems.
This is because directly routing from input sequences to output sequences is problematic in the perspectives of both memory consumption and computational complexity.
For the routing mechanism, prediction vectors representing all the paths from the lower-level capsules to the higher-level capsules are given as inputs, then each of them is weighted by the routing coefficient.
Thus, the size of the transformation matrix to construct the vectors and the number of routing coefficients exponentially increase according to the length of input and output sequences.
As a result, the memory requirements can exceed the acceptable size.
Moreover, the computational burden resulting from the vector construction and the routing method can interfere with the real-time processing of ASR systems.

In this paper, we introduce the sequential routing framework (SRF) which is a novel method to build up CTC networks based on a CapsNet-only architecture.
SRF is applicable to iterative routing-by-agreement methods \citep{DBLP:conf/nips/SabourFH17, DBLP:conf/iclr/HintonSF18} which update the routing coefficients of the current iteration using the outputs of the previous iteration in an expectation and maximization manner.
We chose to apply dynamic routing (DR)~\citep{DBLP:conf/nips/SabourFH17} to the framework since it has shown competitive accuracy and intuitive mechanisms compared to more recent architectures \citep{DBLP:conf/nips/HahnPK19, Malmgren1314210}.
To train SRF models, the input sequences are first converted to three-dimensional sequences through convolutional and linear projection layers.
The encoded sequences are sliced by time windows and multiple routing iterations are performed for each slice to classify the corresponding label.
By slicing capsule groups, the existing routing mechanisms for the fixed size data can be applied.
In addition, the models are trained to use the limited context when encoding each frame, thus the proposed method can have online processing capabilities unlike the architectures that require a full-sequence as an input such as bidirectional LSTMs (BLSTMs) and Transformers.
Afterwards, the training loss is calculated using CTC.
The framework achieved competitive accuracy while minimizing decoding speed degradation and required parameters by sharing two types of information during routing across the slices.
First, the transformation matrices are shared so that only the fixed size of parameters is required regardless of input lengths.
Moreover, the capsule clustering information is also conveyed to the next slice by initializing routing coefficients of the current slice based on the previous routing results.
As a result, only one routing iteration is required for each slice meanwhile the routing coefficients are updated by the number of times corresponding to the each slice index.

The clear contribution of this study is that, to the best of our knowledge, SRF is the first CapsNets only architecture for seq2seq speech recognition. 
The proposed method achieved competitive performance on speech recognition in terms of accuracy and online processing capability by sharing the learnable weights and the clustering information.

\section{Preliminaries}

\subsection{CTC loss function}
A CTC network \citep{DBLP:conf/icml/GravesFGS06} is defined as a continuous map $\mathcal{N}_\theta: (\mathbb{R}^F)^T \mapsto (\mathbb{R}^{V})^T$ from an $F$ dimensional input sequence $x$ of length $T$ to the same length sequence $\hat{y}$ of $V$ dimensional probability vectors with parameters $\theta$.
$V$ is the cardinality of an expanded label set $\mathbb{L}^\prime$ consisting of the union between the label symbol set $\mathbb{L}$ and a blank symbol, $\mathbb{L} \cup \{<$blank$>\}$.
This network contains a softmax layer at the top of it in order to convert logits to valid probability distributions.
The softmax layer for a given vector $Z \in \mathbb{R}^D$ is computed as follows:
\begin{equation}
\text{softmax}(Z)_i = \frac {\exp(z_i)}{\sum^D_{j=1} \exp(z_{j})}, \text{for } i=1,...,D
\label{eq:softmax}
\end{equation}
, where $z_i$ is the $i$-th element of $Z$.
CTC computes a conditional probability of a label sequence $y \in \mathbb{L} ^{\leq T}$ for a given input sequence $x$ by summing up every conditional probability of possible paths $\pi$, i.e., $p(y|x) = \sum_{\pi \in \mathcal{B}^{-1}(y)} p(\pi|x)$.
The possible paths are computed using an inverse of a map $\mathcal{B}: \mathbb{L}^{\prime T} \mapsto \mathbb{L}^{\leq T}$ from an expanded label sequence $y^\prime$ to $y$.
$\mathcal{B}$ performs many-to-one mappings by simply removing repeating and blank symbols from the given paths.
For example, $\mathcal{B}$(``cc-aaa-tt") = $\mathcal{B}$(``-cc-aattt") = ``cat", where ``-" indicates a blank symbol.
It allows CTC networks to learn alignments solely from the input and output sequence pairs.
A conditional probability of each possible path is computed as $p(\pi|x)=\prod_{t=1}^T p(y^{\prime t}_\pi |x^t)$, where $y^{\prime t}_\pi$ is an expanded label symbol in a path $\pi$ at time $t$ and $x^t$ is the $t$-th feature vector of $x$.
The CTC loss is defined as a negative of the summation of the log probabilities of all CTC paths.
The gradients are computed by differentiating the probability function $p(y|x)$ using the CTC forward-backward algorithm, which is a kind of dynamic programming.

\subsection{Capsule Network}
\begin{figure}[tb]
  \includegraphics[trim={0 0cm 0 0cm}, clip, width=1.0 \linewidth]{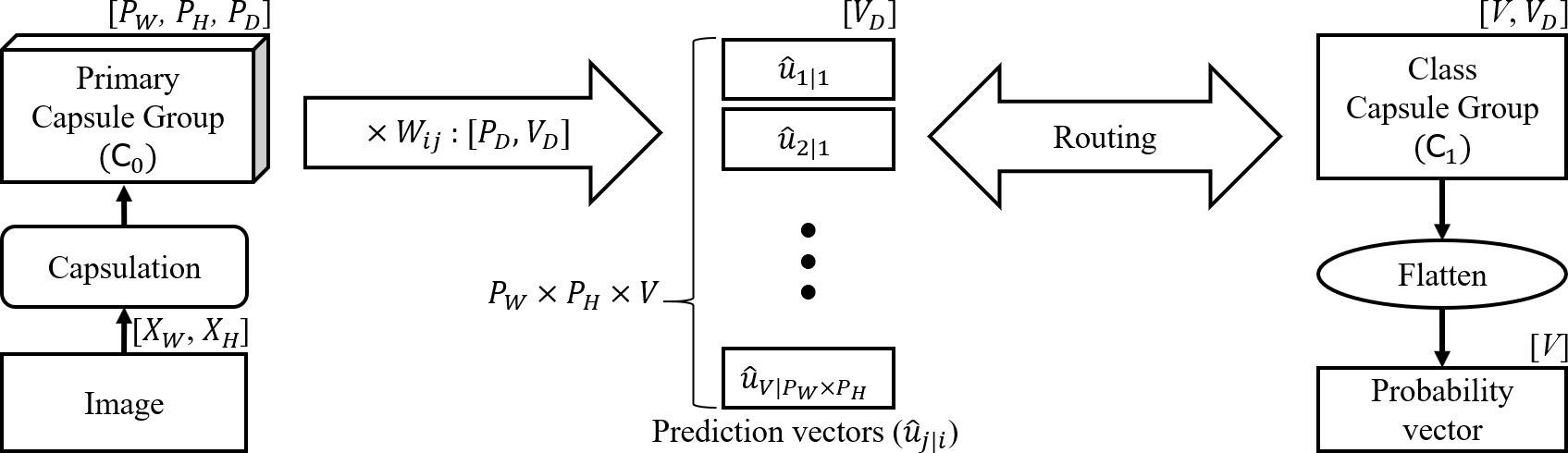}
  \caption{A diagram of a capsule network (CapsNet) consisting of one capsule layer. ($i\in [1, P_W \times P_H]$ and $j \in [1, V]$)}
  \label{fig:capsnet}
\end{figure}
A CapsNet~\citep{DBLP:conf/icann/HintonKW11, DBLP:conf/nips/SabourFH17, DBLP:conf/iclr/HintonSF18} for a two-dimensional data classification problem is defined with the parameter $\psi$ as a map, $\mathcal{C}_\psi: (\mathbb{R}^{X_W})^{X_H} \mapsto \mathbb{R}^{V}$, which maps an input data of size $X_W \times X_H$ to a $V$-dimensional probability vector, e.g., $X_W$ and $X_H$ represent the width and height of the two-dimensional input images respectively as shown in Fig~\ref{fig:capsnet}.
In the figure, the dimension above each capsule group block indicates the dimension of a group of instantiation parameter vectors.
To initiate the routing method, the input is converted to a capsule group, $\mathsf{C}=\{\mathsf{A}, \mathsf{U}\}$.
In the capsulation process, $X_W$ and $X_H$ are converted to the width $P_W$ and the height $P_H$ of the lowest capsule group respectively. Thus, $\mathsf{C}$ consists of a pair of an activation group, $\mathsf{A} \in \mathbb{R}^{P_W \times P_H}$, and a group of instantiation parameter vectors, $\mathsf{U} \in \mathbb{R}^{P_W \times P_H \times P_D}$, where $P_D$ is the dimension of parameter vectors of capsules in the lowest level.
$\mathsf{A}$ can be computed either from $\mathsf{U}$~\citep{DBLP:conf/nips/SabourFH17} or from additional neural layers~\citep{DBLP:conf/iclr/HintonSF18}.
Routing iterations are performed between the higher-level capsules and the prediction vectors $\hat{u_{j|i}}$, each of which represents a path from the $i$-th lower-level capsule to the $j$-th higher-level capsule. Thus, in Fig~\ref{fig:capsnet}, the range of $i$ and $j$ are $[1, P_W \times P_H]$ and  $[1, V]$ respectively. $\hat{u_{j|i}}$ is computed by multiplying an instantiation parameter of the $i$-th lower-level capsule with a transformation matrix $W_{ij}$.
During the training process, for a certain entity type, each of the elements of $\mathsf{A}$ and $\mathsf{U}$ can be trained to represent an existence probability and characteristics respectively~\citep{DBLP:conf/icann/HintonKW11}. Accordingly, CapsNets can potentially~\citep{NIPS2018_8100} represent invariance of the existence probabilities and also equivariance of the properties of the entity type.

The groups in the lowest, highest, and in-between levels are called as primary, class, and convolutional capsule groups respectively.
In this paper, for the sake of brevity, we describe the structure of $\mathsf{U}$ as the structure of the capsule group because the structure of $\mathsf{A}$ can be derived by removing the $D$ dimension from the structure of $\mathsf{U}$.
The first and second dimensions of the class capsule groups are $V$ and $V_D$ respectively, and their activation groups are the outputs of the CapsNets, i.e., the CapsNets output $V$-dimensional vectors.
Gradient based learning methods used for conventional neural networks are applicable.
A key procedural difference of CapsNets is that they filter information flow using routing-by-agreement.

\subsubsection{Dynamic Routing}
DR~\citep{DBLP:conf/nips/SabourFH17} is an iterative routing-by-agreement method which works in a non-parametric expectation and maximization manner based on the similarities between capsules.
In this section, in order to distinguish instantiation vectors in the lower- and higher-level, we use $u$ and $o$ respectively. We also use $i$ and $j$ to represent the index of the vectors in lower- and higher-levels respectively.
The $j$-th activation scalar $a_j$ is computed by the length of the $j$-th instantiation vector $o_{j}$ as follows:
\begin{equation}
a_j = \text{length}(o_j) = \sqrt{\sum^D_{d=1}{o^2_{j_d}}}
\label{eq:length}
\end{equation}
In order to normalize $a_j$ to a valid probability, a nonlinear function, which is referred to as a squash function, is defined as follows:
\begin{equation}
o_j = \text{squash}(s_j) = \frac{||s_j||^2}{1+||s_j||^2} \frac{s_j}{||s_j||}
\label{eq:squash}
\end{equation}
, where $s_j$ is a unnormalized instantiation parameter vector for the $j$-th capsule in the higher-level.
This function also has a role to make the activations more discriminative by pushing most of the values to around zero or one.
A prediction vector, $\hat{u}_{j|i}$, is computed as $\hat{u}_{j|i}=W_{ij} \times u_i$.
The vector indicates a path from the $i$-th capsule in the lower level to the $j$-th capsule in the higher level.
A transformation matrix, $W_{ij}$, is the only required parameter matrix for the routing mechanism.
A $\hat{u}_{j|i}$, iteration number $\Lambda$ and level index $l$ are given for DR.
$l$ is ranged from 0 to $L$ for a CapsNet consisting of $L$ layers, i.e., $l$ of primary capsules is 0.
The primary capsules are calculated through multiple convolutional layers.
Routing coefficients $r$ are zero-initialized, i.e., coupling coefficients $c$ are uniformly initialized.
Then $r$ is updated to maximize the agreements between $\hat{u}_{j|i}$ and $o_j$ as in Algorithm~\ref{algorithm:dr}.
\begin{algorithm}[H]
\small
\caption{Dynamic Routing (DR)~ \citep{DBLP:conf/nips/SabourFH17} (Expectation: Line 7, Maximization: Line 5)}
\begin{algorithmic}[1]
\Procedure{Dynamic Routing}{$\hat{u}_{j|i}, \Lambda, l$}
    \State for all capsule $i$ in level $l$ and capsule $j$ in level $(l+1)$: $r_{ij} \gets 0$
    \For{$\Lambda$ iterations}
        \State for all capsule $i$ in level $l$: $c_i \gets$ softmax($r_i$) \Comment{eq~\ref{eq:softmax}}
        \State for all capsule $j$ in level $(l+1)$: $s_j \gets \sum_i c_{ij} \hat{u}_{j|i}$
        \State for all capsule $j$ in level $(l+1)$: $o_j \gets $ squash$(s_j)$ \Comment{eq~\ref{eq:squash}}
		\State for all capsule $i$ in level $l$ and capsule $j$ in level $(l+1)$: $r_{ij} \gets r_{ij} + \hat{u}_{j|i} \cdot o_j$
    \EndFor
	\State return $o$
\EndProcedure
\end{algorithmic}
\label{algorithm:dr}
\end{algorithm}

\section{Sequential Routing Framework}

\begin{figure}[tb]
  \includegraphics[trim={0 0cm 0 0cm}, clip, width=1.0 \linewidth]{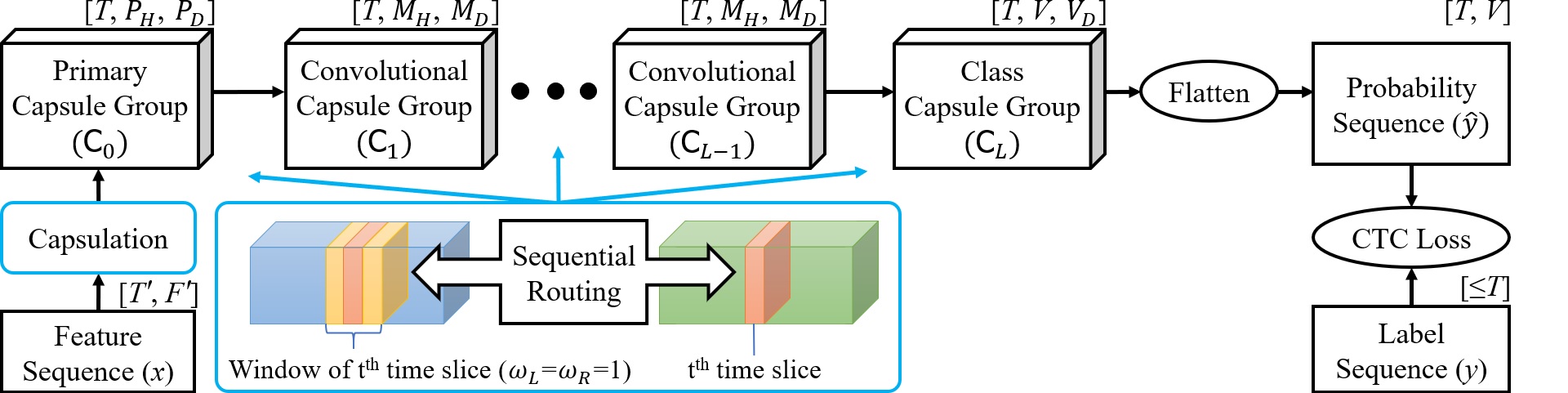}
  \caption{An overview of the sequential routing framework (SRF)}
  \label{fig:diagram}
\end{figure}

SRF is an iterative routing framework for sequence data and it is defined as a modified version of the original CapsNets with the parameter $\varphi$, $\mathcal{S}_\varphi: (\mathbb{R}^{F^\prime})^{T^\prime} \mapsto (\mathbb{R}^V)^T$, from a real valued feature sequence $x$ having a length $T^\prime$ and a feature dimension $F^\prime$ to a $V$ dimensional probability vector sequence $\hat{y}$ of a width $T$ as shown in Fig.~\ref{fig:diagram}.
The shapes of an input, output sequence and each capsule group are written above the boxes in the diagram.
A two-dimensional input feature sequence $x$ is transformed into a primary capsule group $\mathsf{C}_{0}$, whose width, height and depth are $T$, $P_H$ and $P_D$ respectively, through a capsulation block.
Afterwards, $\mathsf{C}_{0}$ is fed into the lowest capsule layer, then encoded to a convolutional capsule group $\mathsf{C}_{1}$, whose width, height and depth are $T$, $M_H$, and $M_D$ respectively.
For the sake of brevity, we describe all the convolutional capsule groups $\mathsf{C}_{1..L-1}$ have the same shape.
A class capsule group $\mathsf{C}_{L}$ has $T$, $V$, and $V_D$ as width, height and depth respectively.
It is flattened to an activation vector sequence, i.e., $\hat{y}=\mathsf{A}_{L}$, which is a sequence of probability vectors where each element indicates a probability of observing a corresponding label symbol.
Finally, the CTC loss between $\hat{y}$ and a label sequence $y$ is computed.

\subsection{Capsulation}

Capsulation is a neural layer block that converts from $x$ to $\mathsf{C}_{0}$, as shown in Fig.~\ref{fig:capsulation}.
$\mathsf{A}$ has the structure of width $T$ and height $P_H$ and the structure of $\mathsf{U}$ has the $P_D$ dimension in addition to the two dimensions of $\mathsf{A}$.
The layers to compute $\mathsf{A}$ are optional structures depending on the routing mechanism.
$x$ is first encoded into three-dimensional representation with width $T$, height $F$ and depth $F_D$ through $L_C$ two-dimensional convolutional layers, where $T \leq T^\prime$ and $F \leq F^\prime$.
We employed maxout~\citep{DBLP:journals/corr/abs-1302-4389} as activation functions of the convolutional layers because of their competitive accuracy in ASR~\citep{DBLP:conf/interspeech/ZhangPBZLBC16}.
Thus, the $l_c$-th convolutional layer in the first sub-block is defined as follows:
\begin{figure}[t]
  \centering
  \includegraphics[trim={0 0cm 0 0cm}, clip, width=1.0\linewidth]{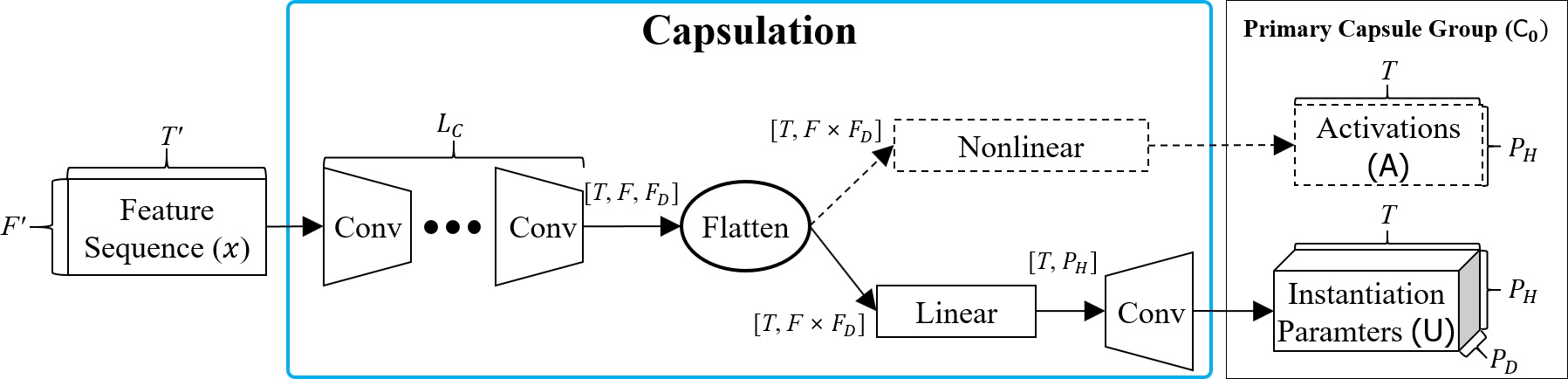}
  \caption{A capsulation block (Dashed boxes are optional blocks)}
  \label{fig:capsulation}
\end{figure}
\begin{equation}
x_{l_c}=\max_{n_c \in [1, N_C]} \text{Dropout}_{n_c}([W_{l_c, n_c, f_d}*x_{l_c-1}]_{f_d=1}^{F_D}, \alpha_d)\text{, }l_c \in [1, L_C]
\label{eq:maxout}
\end{equation}
, where $*$ is a convolutional operator, $N_C$ is the number of convolutional operations in a layer and $\alpha_d$ is the dropout rate. $W_{l_c, n_c, f_d}$ indicates a convolution weight matrix for the $f_d$-th channel of the $n_c$-th convolutional operation in the $l_c$-th layer.
The $0$-th $x$ is an input sequence, i.e., $x_0 \in \mathbb{R}^{F^\prime \times T^\prime}$.
For all the cases, we set $N_C$ and $\alpha_d$ to 2 and 0.2 respectively.
Subsequently, the output sequence is flattened to $x^\prime_{L_C}$ by reshaping it to width $T$ and height $F \times F_d$, then it is fed into two different layer blocks.
They are projected to a normalized vector sequence, i.e., an activation group $\mathsf{A}$, with width $T$ and height $P_H$, through a nonlinear layer as follows:
\begin{equation}
\mathsf{A}=g(W_\mathsf{A} \times x^\prime_{L_C})
\label{eq:nonlinear}
\end{equation}
, where $W_\mathsf{A}$ is a learnable weight matrix and $g$ is a nonlinear function for normalizing the values.
They are also projected to another representation $\mathsf{U}^\prime$ having the same shape with $\mathsf{A}$ as follows:
\begin{equation}
\mathsf{U}^\prime=W_\mathsf{U} \times x^\prime_{L_C}
\label{eq:linear}
\end{equation}
, where $W_\mathsf{U}$ is a learnable weight matrix.
Afterwards, $\mathsf{U}^\prime$ is converted to $\mathsf{U}$ by expanding their channel dimensions into $P_D$ through a two-dimensional convolutional layer activated with the maxout function as follows:
\begin{equation}
\mathsf{U}=\max_{n_c\in[1, N_C]}\text{Dropout}_{n_c}([W_{n_c, p_d} * \mathsf{U}^\prime]_{p_{d}=1}^{P_D}, \alpha_d)
\label{eq:maxout2}
\end{equation}
, where $W_{n_c, p_d}$ indicates a convolutional weight matrix for the $p_d$-th output channel of the $n_c$-th convolutional operation.

\subsection{Routing-by-agreement for sequence to sequence learning}
In SRF, in order to adapt the existing routing methods with minimal changes and to control the required size of parameters regardless of the length of input sequences, each subgroup of a capsule group, $\mathsf{C}=\{\mathsf{A}, \mathsf{U}\}$ is sliced by $T$ without overlapping adjacent slices, i.e., $\mathsf{C}^t=\{\mathsf{A}^t, \mathsf{U}^t\}$.
In order to expand a receptive field on $\mathsf{C}_{0}$, sequential routing is performed between windows of the time slices in the lower level, which consist of the consecutive $\omega$ slices, and single slices in the higher level.
The bottom box in Fig.~\ref{fig:diagram} describes an example of sequential routing between the window ($\omega = 3$) centered on $t$-th slice in the lower level and the $t$-th slice in the higher level.
In this study, we set the stride of the sliding window to one in all the cases and the both side of window contexts beyond the sequence boundaries are padded as zero.
Thus when $\Lambda$ is set to 1, the routing iteration is performed $T$ times for a capsule group having width $T$.
The receptive field on $\mathsf{C}_{0}$ is computed as $\omega + (L-1) \times (\omega - 1)$ for each slice of $\mathsf{C}_{L}$.
Accordingly, online decoding is allowed with a time delay corresponding to $L \times \omega_R$, where $\omega_R$ is the right context size of the window.
In each capsule layer, the $t$-th prediction vector $\hat{u}_{j|i}^t$ is calculated from the $t$-th instantiation vector $u^t$ through a linear transformation using $W_{ij}$ as follows:
\begin{equation}
\hat{u}_{j|i}^t = W_{ij} \times u_i^t, i \in [1, \omega \times I_H], j \in [1, O_H]
\label{eq:uhat}
\end{equation}
, where $I_H$ and $O_H$ are heights of lower and higher capsule groups respectively.
$I_H$ of the first layer and $O_H$ of the last layer are the $P_H$ and $V$ respectively. The two heights are $M_H$ in-between layers.
Accordingly, each $W_{ij}$ has the shape of the product of depths of lower and higher capsule groups.
$W_{ij}$ are shared across all the time slices.
Therefore, the number of parameters for the routing mechanism in a layer is controlled only by the shape of capsule groups and $\omega$.

\begin{figure}[tb]
\centering
\subfloat[][Original routing]{
\includegraphics[width=0.266\linewidth]{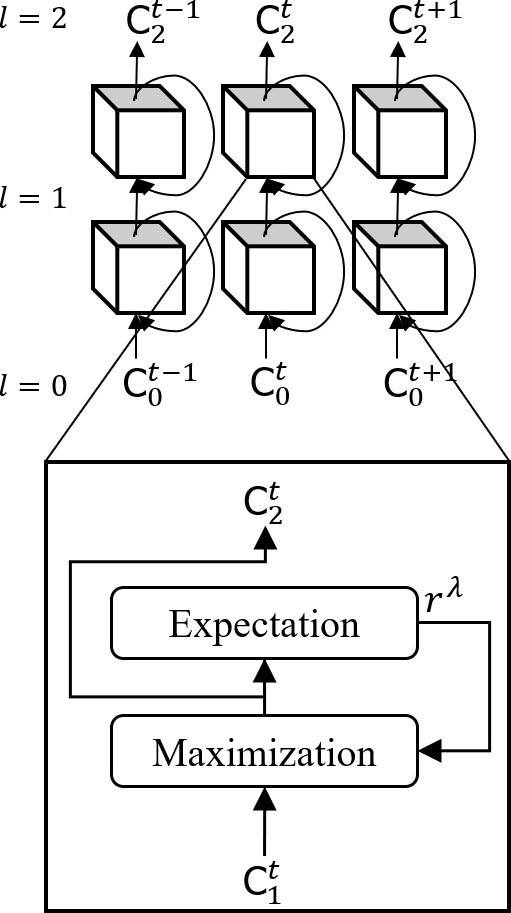}
\label{fig:diagram_3_or}
}
\qquad\qquad
\subfloat[][Sequential routing]{
\includegraphics[width=0.34\linewidth]{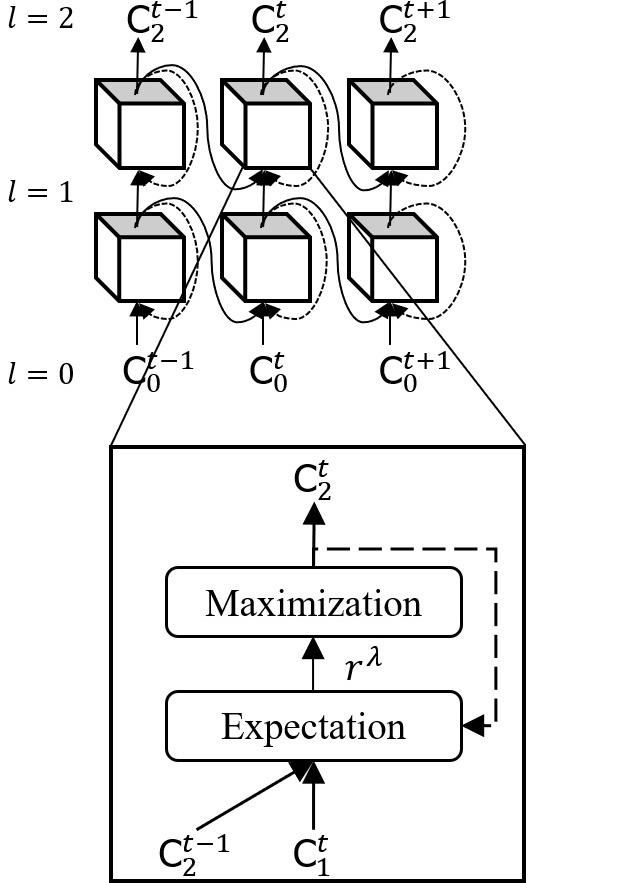}
\label{fig:diagram_3_sr}
}
\caption{Schematic diagrams of the two versions of iterative routing procedures ($L=2$, $\omega=1$, $\lambda \in [1, \Lambda]$, $\dashrightarrow$ and $\rightarrow$ are optional and required procedures respectively)}
\label{fig:diagram3}
\end{figure}

We have an assumption that consecutive slices in sequence data have similar properties.
Based on that, we designed an iterative routing mechanism which initializes routing coefficients for the $t$-th slice based on the routing output of the $(t-1)$-th slice.
Fig.~\ref{fig:diagram3} is a schematic diagram of the two different iterative routing mechanisms of $L=2$ and $\omega=1$ for three consecutive slices.
In this figure, the three dimensional block represents the routing procedures, $\lambda$ is an iteration index in the range between 1 and $\Lambda$, and the dashed and solid arrows describe the optional and required flows respectively.
In the original version of routing, $r$ is initialized uniformly, then updated within each $t$ iteratively as in Fig.~\ref{fig:diagram_3_or} in the order of a maximization and an expectation stage.
The expectation stage means to update $r$ to improve the agreements between adjacent capsule groups and the updated coefficients are applied in the maximization stage.
Thus, if $\Lambda$ is set to 1, output capsule groups are computed using the uniform $r$ in the original routing method.
The proposed sequential routing method has two procedural differences as described in Fig.~\ref{fig:diagram_3_sr}.
First, $r$ is initialized based on the agreement between the routing output of the previous time slice $\mathsf{C}_{2}^{t-1}$ and the current routing input $\mathsf{C}_{1}^{t}$ thus $r$ is uniformly initialized only at $t=1$.
The second procedural difference is the sub-procedure order of the routing mechanism which expects the initial $r$ before the first maximization stage.
These two modifications make a similar effect of updating $r$ $(t-1)$ times to compute $\mathsf{C}^t$ when $\Lambda=1$.
In other words, they alleviate the need for iterative updates of $r$ within each slice as an option.
Consequently, decoding can be performed in a non-iterative way while minimizing accuracy degradation.

\begin{algorithm}[t!]
\small
\caption{Sequential version of Dynamic Routing (SDR) algorithm (Expectation: Line 7, Maximization: Line 9)}
\begin{algorithmic}[1]
\Procedure{Sequential Dynamic Routing}{$o^{t-1}, \hat{u}_{j|i}^{t}, \Lambda, l$}
    \State for all capsule $i$ in $t$-th window of level $l$ 
    \State \hskip1.0em and capsule $j$ in $t$-th slice of level $(l+1)$: $r_{ij} \gets 0$
    \State for all capsule $j$ in $t$-th slice of level $(l+1)$: $o_{j}^{t} \gets o_{j}^{t-1}$
    \For{$\Lambda$ iterations}
        \State for all capsule $i$ in $t$-th window of level $l$ 
        \State \hskip1.0em and capsule $j$ in $t$-th slice of level $(l+1)$:
        $r_{ij} \gets r_{ij} + \hat{u}_{j|i}^t \cdot o_{j}^{t}$
        \State for all capsule $i$ in $t$-th window of level $l$: $c_i \gets$ softmax($r_i$)\Comment{eq~\ref{eq:softmax}}
        \State for all capsule $j$ in $t$-th slice of level $(l+1)$: $s_j \gets \sum_i c_{ij} \hat{u}_{j|i}^t$
        \State for all capsule $j$ in $t$-th slice of level $(l+1)$: $o_{j}^t \gets $ squash$(s_j)$\Comment{eq~\ref{eq:squash}}
    \EndFor
    \State return $o^t$
\EndProcedure
\end{algorithmic}
\label{algorithm:sdr}
\end{algorithm}

Sequential version of DR (SDR) works as Algorithm~\ref{algorithm:sdr} for each $t$ between the $l$-th and $(l+1)$-th level.
To explain the algorithm, we use the notations $u$, $o$, and their indices $i$ and $j$ respectively as explained in Section 2.2.1.
$o^{t-1}$ is zero-initialized at $t=1$.
The expectation-maximization clustering is performed from line 5 to 11.
At line 7, $r_{ij}$ is updated by accumulating the agreements between $\hat{u}_{j|i}^t$ and $o_{j}^{t}$.
To maximize the agreements, $s_j$ is computed as the summation of $\hat{u}_{j|i}^t$ over all $i$ by weighting with the updated $c_{ij}$ at line 9.
The number of operations of an iteration remains the same as Algorithm~\ref{algorithm:dr} since only the order of the sub-procedures is changed.

\section{Results}
All evaluations were performed with the same settings when it comes to training CapsNets unless otherwise noted.
Variables are initialized using a fan-avg method~\citep{DBLP:journals/jmlr/GlorotB10} from the uniform distribution, i.e., learning weights are drawn from $[-\sqrt{3 \times \alpha_s / n_{init}}, \sqrt{3 \times \alpha_s / n_{init}}]$, where $n_{init}$ is the average of the input and output unit numbers and the scaling factor $\alpha_s$ is set to 1.0.
For all SRF models, dropout~\citep{DBLP:journals/jmlr/SrivastavaHKSS14} layers are applied after every layer at rate 0.2.
We utilized two types of normalization layers.
First, batch normalization~\citep{DBLP:conf/icml/IoffeS15} layers are added after every convolutional layer in the first convolutional sub-block of a capsulation block.
Second, layer normalization~\citep{DBLP:journals/corr/BaKH16} layers are also applied between every capsule layer.
One modification is that layer normalization is performed not for each capsule but over all capsules in the same time slice.
An Adam optimizer~\citep{DBLP:journals/corr/KingmaB14} is used for the gradient descent algorithm.
A learning rate is updated for each step $n_{s}$ depending on the two hyper-parameters which are a warming-up step $n_{w}$ and a scaling factor $\kappa$ as follows:
\begin{equation}
\text{Learning Rate} = \kappa \times min(n_{s}^{-0.5}, n_{s} \times n_{w}^{-1.5})
\label{eq:lr}
\end{equation}
The beam size for decoding is set to 100.
The proposed method was implemented with Tensorflow~\citep{DBLP:journals/corr/AbadiABBCCCDDDG16}.
The error rates were evaluated with SCTK\footnote{\url{https://github.com/usnistgov/SCTK}}.

\subsection{The TIMIT Corpus}
TIMIT~\citep{timit} consists of mono-channel read speech sampled at 16Khz.
The training and test set consist of 4,620 utterances recorded from 462 speakers and 1,680 utterances recorded from 168 speakers respectively.
We used a training set consisting of 3,696 utterances where all dialect utterances, i.e., the utterances tagged as ``SA", were removed and used 192 sentences from the core test set recorded from 24 speakers.
A validation set was selected from another portion of the test set and was made up of 400 utterances recorded from 50 different speakers.
A total of 63 labels consisting of 61 phonemes plus a padding and blank symbol were used during both training and decoding.
At the top layer, the values in $c$ which route to a class capsule corresponding to the padding symbol are masked as zero.
To evaluate phoneme error rates (PERs), the phoneme labels were mapped to 39 labels~\citep{DBLP:journals/tsp/LeeH89}.
The features were extracted with a 10 ms hop size and 25 ms window size, and were encoded with 40-dimensional Fourier-transform-based filterbanks plus energy. Their temporal first and second-order differences were added with the delta-window size 2, thus 123-dimensional vectors were used as inputs.
The input features were normalized to zero mean and unit variance per-speaker.
The data splitting and feature extraction were performed using Kaldi\footnote{\url{https://github.com/kaldi-asr/kaldi}}\citep{Povey_ASRU2011}.

The learning schedule, $n_{w}$ is set to 1,200.
We applied an additional decay policy where the learning rate was started by setting $\kappa$ to 0.5, and then $\kappa$ was reduced to 0.1 after 27 epochs.
The models were trained for 200 epochs to ensure sufficient weight updates.
In order to avoid the accuracy being dependent on the early stop time, we evaluated PERs with a model which is the averaged checkpoint of the last 10 epochs.
Approximately 5K frames are contained in a batch for each training step according to their sequence length thus the learnable weight are updated 42K times per experiment.
We first investigated the performance gain from the proposed routing algorithm using small CapsNet models with $L=1$ and $P_H=20$.
\begin{table}[th!]
\centering
\begin{tabular}{cccc}
Routing & \multirow{2}{*}{Iteration} & \multicolumn{2}{c}{PER(\%)} \\
Method  &                            & Valid & Test                \\ \hline
\multirow{3}{*}{DR}  & 1 & 25.6 \footnotesize{$\pm0.4$} & 26.8 \footnotesize{$\pm0.3$} \\
                     & 2 & 25.4 \footnotesize{$\pm0.2$} & 26.6 \footnotesize{$\pm0.4$} \\
                     & 3 & 25.4 \footnotesize{$\pm0.3$} & 26.8 \footnotesize{$\pm0.4$} \\ \hline
\multirow{3}{*}{SDR} & 1 & 24.4 \footnotesize{$\pm0.4$} & 25.5 \footnotesize{$\pm0.4$} \\
                     & 2 & 24.3 \footnotesize{$\pm0.5$} & 25.7 \footnotesize{$\pm0.3$} \\
                     & 3 & 24.5 \footnotesize{$\pm0.3$} & 25.8 \footnotesize{$\pm0.3$}
\end{tabular}
\caption{Phone error rates (PERs) according to the routing methods and iteration numbers. The means and 95\% confidence intervals of PERs were obtained from a total of 5 experiments.}
\label{Tab:iter}
\end{table}

We compared SDR with DR as in Table~\ref{Tab:iter}.
$\omega$ and the depth of capsule groups are set to 1 and 8 respectively.
Accordingly, the total number of parameters of each model is 207,682.
In each of the two cases, statistically significant differences in PERs depending on the number of iterations are not observed.
However, in all the cases, SDR shows about 1\% lower PERs than that of DR.
These experiments will be further analyzed in Section 5 by investigating the heat maps of $c$.

\begin{table}[b!]
\centering
\begin{tabular}{cc|ccc|cccc}
\multicolumn{2}{c|}{Window ($\omega$)}&Look-ahead &Delay  & Params. &\multicolumn{2}{c}{Valid(\%)}&\multicolumn{2}{c}{Test(\%)} \\ 
$\omega_L$ & $\omega_R$ & frames & (ms)  & (M)  & PER  & EOS  & PER  & EOS   \\\hline
0          & 0          & 11     & 122.5 & 0.21 & 24.4 & 52.8 & 25.6 &  55.7 \\
1          & 0          & 11     & 122.5 & 0.30 & 23.8 & 25.5 & 25.1 &  26.6 \\
1          & 1          & 15     & 162.5 & 0.39 & 21.8 & 66.5 & 23.7 &  69.3 \\
2          & 0          & 11     & 122.5 & 0.39 & 23.3 & 27.5 & 24.5 &  26.6 \\
2          & 1          & 15     & 162.5 & 0.48 & 22.5 & 53.0 & 23.6 &  53.1 \\
2          & 2          & 19     & 202.5 & 0.57 & 20.2 & 99.8 & 21.7 & 100.0 \\
3          & 2          & 19     & 202.5 & 0.66 & 20.8 & 94.0 & 21.9 &  93.8 \\
4          & 1          & 15     & 162.5 & 0.66 & 21.1 & 84.5 & 22.6 &  81.3 \\
5          & 0          & 11     & 122.5 & 0.66 & 22.3 & 32.5 & 23.9 &  32.3
\end{tabular}
\caption{Phoneme error rates (PERs) and end-of-sentence (EOS) detection rates depending on the window ($\omega$) configurations.}
\label{Tab:window}
\end{table}

We evaluated SRF models according to the configurations of $\omega$ as in Table~\ref{Tab:window}.
Their $\Lambda$ and capsule group depths are set to 1 and 8 respectively.
As $\omega$ is expanded from 1 to 6, the numbers of required parameters (Params.) are increased from 0.21 to 0.66 million (M) as explained in Section 3.2.
With the consideration of algorithmic delay caused by $\omega_R$, all models are set to have $\omega_L$ either longer than or equal to $\omega_R$.
The time delays were estimated by ``hop size (10ms) $\times$ look-ahead frames + 12.5ms (the half of window size 25ms)". The number of look-ahead frames includes 4 more frames in addition to the frames for the setting of $\omega_R$ due to computing the delta plus double-delta features.
The relative PER reductions (RPERRs) are at most 17.2\% and 15.2\% for Valid and Test respectively depending on $\omega$.
End-of-sentence (EOS) detection rates seem to be one of the reasons why the settings that have unbalanced contexts show worse PERs than settings with balanced contexts since the detection rates show a large range from about 26\% to 100\% for both sets according to the setting of $\omega$.

\begin{figure}[tb!]
\centering
\subfloat[][Phoneme recognition rates (PERs) according to $\omega$]{
\includegraphics[width=0.48\linewidth]{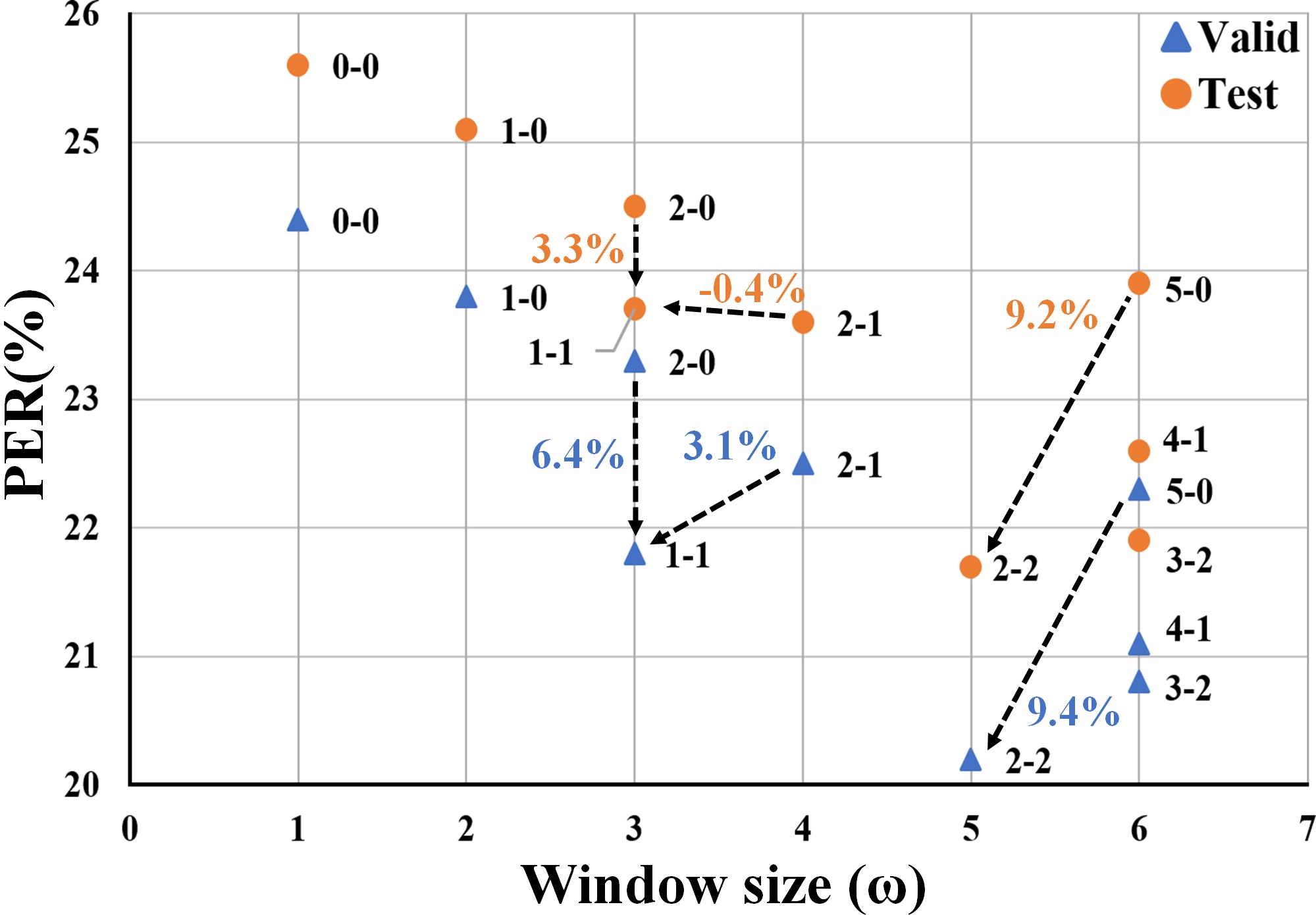}
\label{fig:diagram_win1}
}
\qquad
\subfloat[][Average end-of-sentence (EOS) detection rates on the Valid and Test sets according to $\omega_R$.]{
\includegraphics[width=0.42\linewidth]{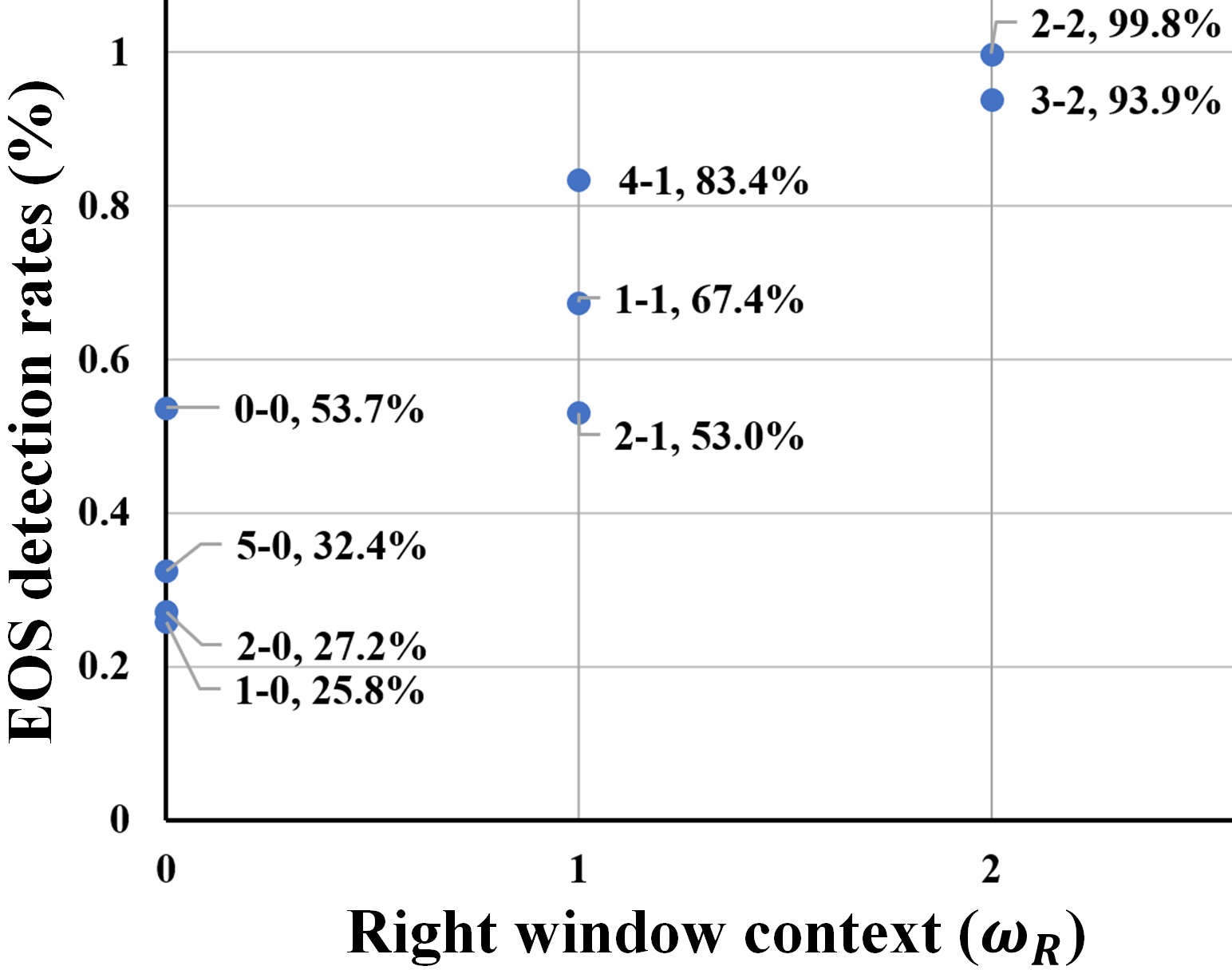}
\label{fig:diagram_win2}
}
\caption{The relationship between performances and the configurations of the window such as the window size $\omega$ and the right context size of the window $\omega_R$.}
\label{fig:diagram_window}
\end{figure}

The relationship between the configuration of the window (X-axis) and PERs (Y-axis) are more explicitly described in Fig.~\ref{fig:diagram_window}.
``2-1" indicates that $\omega_L$ and $\omega_R$ are 2 and 1 respectively.
Blue triangles and orange circles indicate Valid and Test respectively in Fig.~\ref{fig:diagram_win1}.
In the figure, the three evenly sliced cases which are ``0-0", ``1-1" and ``2-2", show almost linear error reductions according to $\omega$ for the both sets.
We can also see multiple cases where the balance of $\omega_L$ and $\omega_R$ has more of an effect on lowering the PERs than the number of parameters when $\omega > 1$.
These phenomena are described by the dashed arrows in the figure and the percentages beside the arrows are RPERRs between the balanced and unbalanced window configurations.
The models ``1-1" and ``2-0" have the same receptive field, but the model ``1-1" shows relatively 6.4\% and 3.3\% lower PERs for Valid and Test respectively.
In addition, although the model ``2-1" has wider receptive field compared to the model ``1-1", it only shows slightly lower PER (relatively 0.4\%) on Test but relatively 3.1\% higher PER on Valid.
Even more, the ``5-0" model shows relatively about 9\% worse PERs than the ``2-2" model on the both evaluation sets.
As $\omega_R$ of models (X-axis) expands, the EOS detection rates (Y-axis) get higher as shown in Fig.~\ref{fig:diagram_win2}.
The detection rates are the averages from the two evaluation sets.
The models which have one longer $\omega_L$ which are ``1-0'', ``2-1'' and ``3-2'', show lower EOS detection rates than that of ``0-0", ``1-1", and ``2-2" respectively.
As the differences between both sides of the window context size get bigger, the PERs and EOS detection rates of the settings get worse, i.e., phone recognition rates (PRRs) and EOS detection rates of the settings show the relationships such as ``5-0” $<$ ``4-1” $<$ ``3-2” with the same number of parameters.

\begin{table}[ht!]
\centering
\begin{tabular}{cccc}
\multirow{2}{*}{Depth} & Params. & \multicolumn{2}{c}{PER(\%)} \\
                       & (M)     & Valid         & Test        \\\hline
 2 & 0.14 & 26.6 & 27.9 \\
 4 & 0.19 & 23.7 & 25.0 \\
 8 & 0.39 & 21.8 & 23.7 \\
16 & 1.15 & 20.6 & 21.9
\end{tabular}
\caption{Phoneme error rates (PERs) depending on the depth of capsule groups}
\label{Tab:depth}
\end{table}

We also evaluated PERs by doubling the capsule group depth continually from 2 to 16, as in Table~\ref{Tab:depth}.
In these experiments, $\omega_L$ and $\omega_R$ are fixed to 1, and $\Lambda$ is set to 1.
The required parameters are increased nearly proportional to the depth of capsules from 0.14M to 1.15M as explained in Section 3.2.
The RPERRs between the depth 2 and 16 cases for Valid and Test are 22.6\% and 21.5\% respectively.
The PER reduction for both sets is 2.9\% when increasing the depth from 2 to 4. Meanwhile, when increasing the depth from 4 to 8, it decreases to less than 2.0\%, even though the increase in the number of parameters is 4 times larger from 0.05M to 0.2M.

\begin{table}[ht!]
\centering
\begin{tabular}{c|ccc|cc}
\multirow{2}{*}{Model} & Look-ahead & Delay       & Params. & \multicolumn{2}{c}{PER(\%)} \\
               & frames         & (ms)            & (M)     & Valid       & Test \\\hline
BLSTM-5L-250H  & Full           & Full            & 6.80    & -           & 18.4 \\
ULSTM-3L-421H  & $\approx$4*    & $\approx$52.5*  & 3.80    & -           & 19.6 \\
CNN-10L-Maxout & $\approx$24*   & $\approx$252.5* & 4.30    & 16.7        & 18.2 \\\hline
TF-5L          & Full           & Full            & 1.99    & 18.9        & 19.9 \\
TF-10L         & Full           & Full            & 3.63    & 18.0        & 19.1 \\
TF-20L         & Full           & Full            & 6.93    & 17.5        & 18.6 \\\hline
SRF-1L         & 15             & 162.5           & 1.01    & 20.1        & 21.5 \\
SRF-2L         & 19             & 202.5           & 0.99    & 18.8        & 20.2 \\
SRF-5L         & 31             & 322.5           & 1.58    & 16.5        & 18.1 \\
SRF-7L         & 39             & 402.5           & 1.97    & 15.9        & 17.5
\end{tabular}
\caption{Phoneme error rates (PERs) of SRF models and other CTC networks such as BLSTM-5L-250H ~\citep{DBLP:conf/icassp/GravesMH13}, ULSTM-3L-421H~\citep{DBLP:conf/icassp/GravesMH13}, CNN-10L-Maxout~\citep{DBLP:conf/interspeech/ZhangPBZLBC16} and Transformer-based CTC networks for the TIMIT corpus~\citep{timit}.
* Calculated by setting the delta-window to 2.}
\label{Tab:pc}
\end{table}

We compared the SRF models with other architectures as in Table~\ref{Tab:pc}. 
The referenced PERs were evaluated using LSTM\citep{DBLP:conf/icassp/GravesMH13} and CNN\citep{DBLP:conf/interspeech/ZhangPBZLBC16}-based CTC networks.
We also compared with Speech-Transformer~\citep{8462506}-based CTC networks which we implemented.
The models have the following structures below. 

\begin{small}
- BLSTM-5L-250H*~\citep{DBLP:conf/icassp/GravesMH13}: BLSTM(250, concat)$\times$5-FC(63)

- ULSTM-3L-421H*~\citep{DBLP:conf/icassp/GravesMH13}: ULSTM(421)$\times$3-FC(63)

- CNN-10L-Maxout*~\citep{DBLP:conf/interspeech/ZhangPBZLBC16}: Conv(3$\times$5, 64)-MP(3$\times$1)-Conv(3$\times$5,

\hspace{0.2cm}64)$\times$3-Conv(3$\times$5, 128)$\times$5-Conv(3$\times$5, 24)-FCMO(512)$\times$2-FCMO(63)

- TF-5/10/20L: CNNFE-FCPE(128)-TF(128, 4, 1024)$\times$L-FC(63)

- SRF-1/2/5/7L: CNNFE-FC(60)-Conv(3$\times$3, 8)-SDR(30, 8)$\times$(L-1)-SDR(63, 8)

* Estimated by considering the number of model parameters.
\end{small}

Each layer is defined as the list below.

\begin{small}
- BLSTM(cell states, merge mode=\{concatenation (concat) or average (ave)\}): a 

\hspace{0.2cm}BLSTM layer

- ULSTM(cell states): a unidirectional LSTM (ULSTM) layer

- Conv(filter size=frequency$\times$frame, output channels, stride=[frequency=1,

\hspace{0.2cm}frame=1]): a two dimensional convolutional layer activated with maxout,

\hspace{0.2cm}the number of feature maps is twice of the channels.

- CNN frontend (CNNFE): Conv(3$\times$3, 64, [2, 2]) $\times$ 2

- MP(pooling size=frequency$\times$frame): a max pooling layer

- FC(output dimension): a fully connected layer

- FCPE(output dimension): a fully connected layer + positional encoding

\hspace{0.2cm}\citep{8462506}

- FCMO(output dimension): a fully connected layer activated with maxout,

\hspace{0.2cm}the number of units is twice of the dimensionality of output space.

- TF(embedding dimension, attention heads, inner layer size): a Transformer 

\hspace{0.2cm}encoder layer~\citep{8462506}

- SDR($O_H$, depth of the output capsule group, [$\omega_L$=1, $\omega_R$=1], $\Lambda$=1): a SDR layer
\end{small}

$\kappa$ is set to 0.13 and reduced to 0.04 under the same condition as in the cases of the CapsNets and $n_{w}$ is set to 1,000.
There are approximately 15K frames in a batch according to the length of utterances.
We set dropout rates for the inputs, attention heads, inner layers and residual connections to 0.3, 0.3, 0.4, and 0.4 respectively.
Bigger penalties are added to non-diagonal elements in attention maps before applying the softmax function Eq.~\ref{eq:softmax}, depending on the distance $\delta$ from the diagonal of the maps in the form of $-\log(1+\delta \times \beta)$ as used in~\citep{8462506}, where the scaling factor $\beta$ were set to 1.0.
TF-5L contains a similar number of parameters as the biggest SRF model SRF-7L.

As the layers of SRF models are stacked up, the receptive fields are increased from 31 to 79 frames.
The reason why SRF-1L requires 0.02M more parameters than SRF-2L is that SRF-1L directly connects primary capsule groups to class capsule groups thus SRF-1L needs 11,340 (60 $\times$ 63 $\times$ 3) transformation matrices while SRF-2L needs 11,070 ((60 $\times$ 30 + 30 $\times$ 63) $\times$ 3).
When comparing the two models, the number of layers seems to be more effective for reducing PERs than the number of parameters.
Among the CTC networks except for SRF models, CNN-10L-Maxout~\citep{DBLP:conf/interspeech/ZhangPBZLBC16} shows the lowest PERs.
Compared to CNN-10L-Maxout, although SRF-5L and SRF-7L require more delays, SRF-5L shows similar PERs with 63.3\% fewer parameters.
Moreover, SRF-7L achieves PERs of 15.9\% and 17.5\% for Valid and Test respectively, which are about 0.7\% lower in PERs for both sets than that of CNN-10L-Maxout, with less than half of the parameters.

\begin{table}[ht!]
\centering
\begin{tabular}{c|c|c|ccc}
\multirow{2}{*}{Model} & Params. & Training time & \multicolumn{3}{c}{Decoding time}  \\
                       & (M)     & (in secs.)    &  Secs. & xRT & $\mathsf{r}$ \\\hline
TF-5L                  & 1.99 &  12 & 12 & 0.02 & 0.90 \\
TF-10L                 & 3.63 &  13 & 13 & 0.02 & 0.93 \\
TF-20L                 & 6.93 &  17 & 16 & 0.03 & 0.93 \\\hline
SRF-1L                 & 1.01 &  98 & 14 & 0.02 & 0.96 \\
SRF-2L                 & 0.99 &  98 & 17 & 0.03 & 0.97 \\
SRF-5L                 & 1.58 & 168 & 25 & 0.04 & 0.98 \\
SRF-7L                 & 1.97 & 215 & 32 & 0.06 & 1.00
\end{tabular}
\caption{Required time to train an epoch of the training set and decode the test set for the TIMIT corpus~\citep{timit}. The correlation coefficients ($\mathsf{r}$) were calculated between SRF-7L and other models.}
\label{Tab:time_timit} 
\end{table}

We measured the training and decoding time of TF-5/10/20L and SRF-1/2/5/7L on a NVIDIA\textsuperscript{TM} RTX-3090 as in Table~\ref{Tab:time_timit}. The training time is in seconds to finish one epoch of training. The size of the train-batch is set to 10K frames so that gradients were updated 107 times. We evaluated the decoding time in seconds and real-time factors (xRT) with Test. SRF-7L requires 18.6 times longer training time compared to TF-5L which has the same number of parameters. The training time of SRF models is increased as the size of models get bigger. On the other hand, decoding time is increased as more layers are stacked up. The correlation coefficients ($\mathsf{r}$) between SRF-7L and other models are larger than 0.90 in all the cases.

\subsection{The Wall Street Journal Corpus}

\begin{table}[b!]
\centering
\begin{tabular}{c|ccc|cc}
\multirow{2}{*}{Model} & Look-ahead & Delay & Params. & \multicolumn{2}{c}{WER(\%)} \\
                       & frame      & (ms)  & (M)  & dev-93 & eval-92 \\\hline
BLSTM-5L               & Full       & Full   & 21.11 & 23.4 & 18.0 \\
ULSTM-5L               & 7          & 82.5   & 21.15 & 33.7 & 27.4 \\
CNN-10L                & 87         & 882.5  & 21.12 & 33.7 & 26.9 \\
CNN-15L                & 127        & 1282.5 & 21.08 & 30.1 & 24.7 \\
TF-20L                 & Full       & Full   & 21.13 & 25.4 & 21.6 \\\hline
SRF-7L-Small           & 67         & 682.5  & 7.75  & 26.4 & 20.3 \\
SRF-7L-Big             & 67         & 682.5  & 15.45 & 25.3 & 19.4 \\
SRF-10L-Small          & 91         & 922.5  & 10.51 & 23.4 & 18.6 \\
SRF-10L-Big            & 91         & 922.5  & 21.13 & 22.4 & 16.9
\end{tabular}
\caption{Word error rates (WERs) of SRF models and other CTC networks for the Wall Street Journal (WSJ) corpus~\citep{wsj0, wsj1}.}
\label{Tab:wsj}
\end{table}

The Wall Street Journal (WSJ) corpus~\citep{wsj0, wsj1} contains mono-channel read speech sampled at 16Khz.
The si284 data set, which consist of about 81-hours of transcribed audio data (37,416 utterances), was set for training.
The dev-93 (503 utterances, 1.1 hours) and eval-92 (333 utterances, 0.7 hours) data set were used respectively to validate and evaluate the models.
For both training and evaluations, a total of 32 labels which consist of 26 uppercase letters, noise marker, apostrophe, space, EOS symbol plus a padding and blank symbol was used.
The feature extraction and training configuration are the same as used in Section 4.1 unless it is specified.
When it comes to the learning schedule, $n_{w}$ was set to 25,000.
$\kappa$ of each model was individually set according to their valid loss curves and was halved at the 71st epoch out of 80 epochs.
There were approximately 20K frames in a batch thus the weights were updated 110K times.
Word error rates (WERs) were evaluated by averaging the last 5\% of checkpoints (4 checkpoints).
As in Table~\ref{Tab:wsj}, the SRF models are compared with other CTC-based ASR systems which we implemented.
The number of parameters of the compared models are set to 21.1M.
The models in Table~\ref{Tab:wsj} were constructed as the list below:

\begin{small}
- BLSTM-5L: CNNFE-BLSTM(441, ave)$\times$5-FC(32)

- ULSTM-5L: CNNFE-ULSTM(662)$\times$5-FC(32)

- CNN-10L: CNNFE-Conv(3$\times$5,140)$\times$4-Conv(3$\times$5, 300)$\times$5-Conv(3$\times$5, 66)-

\hspace{0.2cm}FCMO(1024)$\times$2-FCMO(32)

- CNN-15L: CNNFE-Conv(3$\times$5, 100)$\times$4-Conv(3$\times$5, 215)$\times$10-Conv(3$\times$5, 66)-

\hspace{0.2cm}FCMO(1024)$\times$2-FCMO(32)

- TF-20L: CNNFE-FCPE(256)-TF(256, 4, 1488)$\times$20-FC(32)

- SRF-7/10-Small: CNNFE-FC(52)-Conv(3$\times$3, 16)-SDR(26, 16, [2, 2])$\times$(L-1)-

\hspace{0.2cm}SDR(32, 16, [2, 2])

- SRF-7/10-Big: CNNFE-FC(60)-Conv(3$\times$3, 20)-SDR(30, 20, [2, 2])$\times$(L-1)-

\hspace{0.2cm}SDR(32, 20, [2, 2])
\end{small}

The dropout rates of both LSTM networks are set to 0.3 and 0.4 for input data and in-between layers respectively.
The dropout rates are set to 0.2 between layers for both CNN models.
The layer normalization~\citep{DBLP:journals/corr/BaKH16} layers are added after every layer in both the LSTM and CNN-based CTC networks.
The learning rates for BLSTM-5L and ULSTM-5L were set to 1e-4 and 5e-5 respectively without learning rate scheduling Eq.~\ref{eq:lr}.
The initial $\kappa$ of other models was set to 0.6.
For TF-20L and SRF-7L-Big, it decreased to 0.06 and 0.1, respectively.
Besides the two models, $\kappa$ was initially decreased to 0.5 and then to 0.1.
SRF-10L-Big attains the lowest WER at 16.9\% for eval-92 which is 1.1\% better than that of BLSTM-5L. SRF-7L-Small requires almost half of the delay compared to CNN-15L and shows 7.3\% and 7.1\% lower WERs in dev-93 and eval-92 respectively compared to ULSTM-5L.
SRF-10L-Small shows lower WERs with 68.0\% of the parameters compared to SRF-7L-Big but it requires a 240 ms longer delay.
Compared to TF-20L, relative WER reductions (RWERRs) of SRF-10L-Small are 7.9\% and 13.9\% for dev-93 and eval-92 respectively with half of the parameters.

\begin{table}[t!]
\centering
\begin{tabular}{c|c|c|ccc}
\multirow{2}{*}{Model} & Params. & Training time & \multicolumn{3}{c}{Decoding time}  \\
                       & (M)     & (in secs.)    &  Secs. & xRT & $\mathsf{r}$ \\\hline
BLSTM-5L      & 21.11  & 1,414   & 31   & 0.01 & 0.96 \\
ULSTM-5L      & 21.15  & 1,030   & 30   & 0.01 & 0.96 \\
CNN-10L       & 21.12  & 2,143   & 24   & 0.01 & 0.94 \\
CNN-15L       & 21.08  & 2,310   & 26   & 0.01 & 0.94 \\
TF-20L        & 21.13  & 880     & 29   & 0.01 & 0.91 \\\hline
SRF-7L-Small  &  7.75  & 20,814  & 112  & 0.05 & 0.99 \\
SRF-7L-Big    & 15.45  & 31,406  & 112  & 0.05 & 0.99 \\
SRF-10L-Small & 10.51  & 28,271  & 153  & 0.06 & 0.99 \\
SRF-10L-Big   & 21.13  & 51,666  & 153  & 0.06 & 1.00
\end{tabular}
\caption{Required time to train an epoch of the training set and decode eval-92 for the Wall Street Journal (WSJ) corpus~\citep{wsj0, wsj1}. The correlation coefficients ($\mathsf{r}$) were calculated between SRF-10L-Big and other models.}
\label{Tab:time_wsj}
\end{table}

As in Table~\ref{Tab:time_wsj}, we evaluated the training and decoding time of the models explained in Table~\ref{Tab:wsj} on a NVIDIA\textsuperscript{TM} RTX-3090. The size of train-batch was set to 7K frames thus gradient updates were performed 4,155 times for an epoch of training. The decoding time was measured using eval-92. To finish an epoch of training, SRF-10L-Big takes 59 times longer than TF-20L, which has the shortest training time, and 37 times longer than BLSTM-5L, while showing similar word recognition accuracy. As the parameters and layers of the SRF models increase, their training and decoding time increase respectively. $\mathsf{r}$ between SRF-10L-Big and other models is at least 0.91.

\section{Analysis}
\begin{figure}[ht!]
  \includegraphics[trim={0 0cm 0 0cm}, clip, width=1.0 \linewidth]{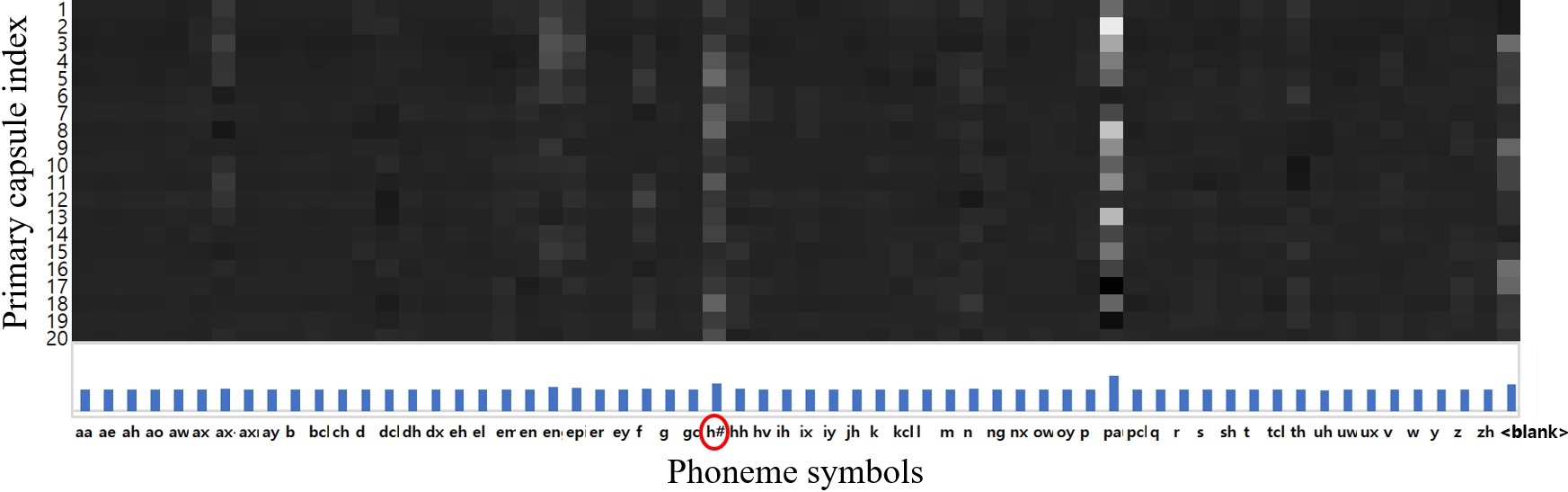}
  \caption{A heat map of coupling coefficients $c$ (SDR, Iteration=1, $t$=137)}
  \label{fig:cc}
\end{figure}
In this section, we investigate how the routing method works in SRF by observing the heat maps of $c$ for a 5.47 seconds length utterance (id: MNJM0\_SI950) in the test set.
In the heat maps, brighter cells refer to larger values ranging from 0.01 to 0.05.
All models explained in this section are the same as explained in Table~\ref{Tab:iter} of Section 4.2.
Fig.~\ref{fig:cc} is a heat map of $c$ which maps from primary capsules (vertical-axis) to class capsules (horizontal-axis) of the iteration 1 version of the SDR model at $t=137$.
A corresponding reference label is the red circled symbol ``h\#”, which indicates the end of a sentence.
The numbers on the vertical-axis indicates primary capsule indexes.
The bar graphs at the bottom of the figure represents the summation of coefficients per each class capsule.
The summation of each row, i.e., the summation of coefficients per each primary capsule, is one and the coefficients which route capsules to the padding symbol are masked to zero as explained in Section 3.2 and they are not represented in the heat map.
As shown in the figure, primary capsules are mostly routed to a class capsule corresponding to a ``pau” symbol with the accumulated coefficient of 0.52 then followed by ``h\#" with that of 0.41 and ``$<$blank$>$" with that of 0.39.
7 primary capsules indexed by in the order of 17, 19, 20, 6, 7, 12 and 5 are routed more to the class capsule corresponding to the correct symbol ``h\#" rather than the capsule for ``pau".

\begin{figure}[ht!]
  \includegraphics[width=1.0 \linewidth]{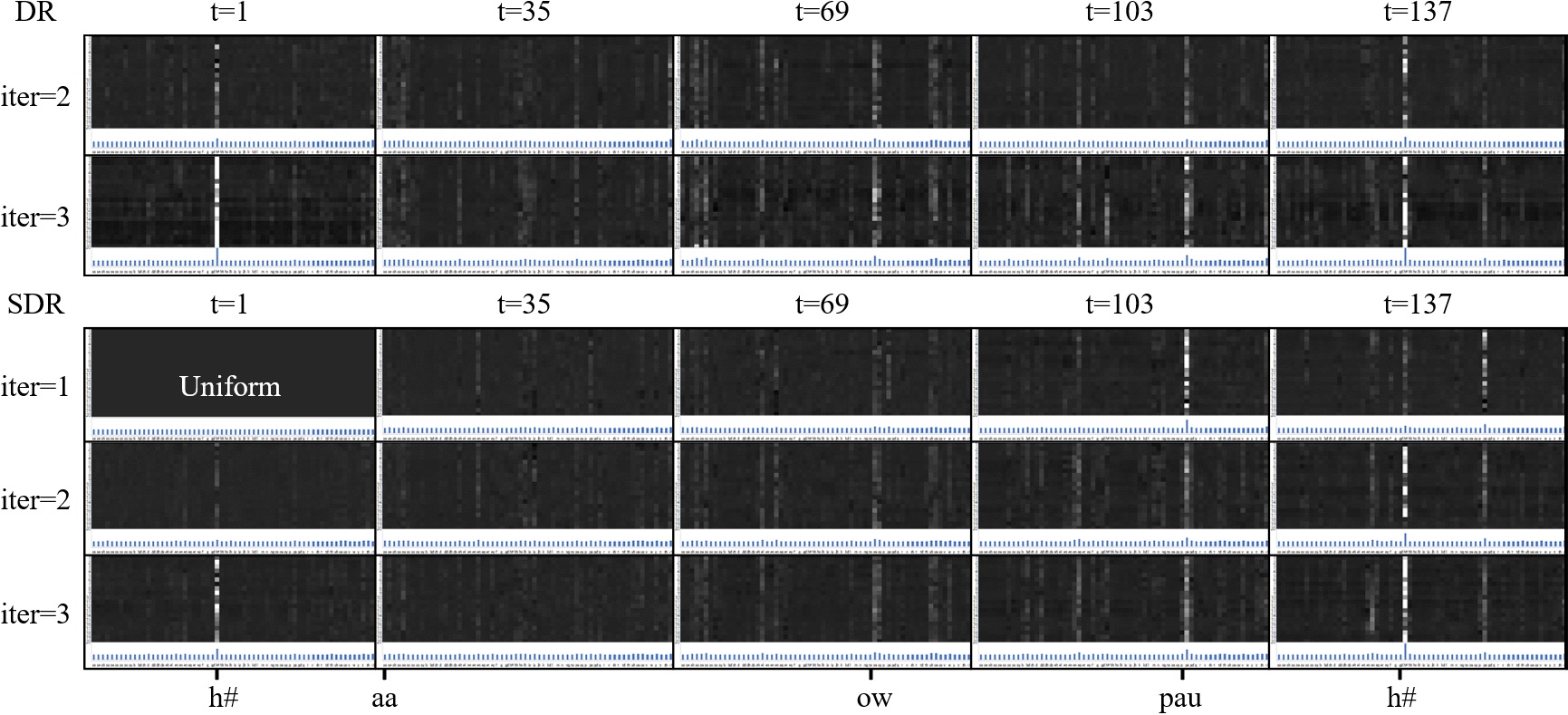}
  \caption{Heat maps of coupling coefficients $c$ for DR and SDR}
  \label{fig:ccm}
\end{figure}

We compare 25 heat maps for different iteration numbers from 1 to 3 between DR and SDR as in Fig.~\ref{fig:ccm}.
The coefficients of iteration 1 version of DR are all the same so it is not included in the figure.
The reference symbols for each $t$ are written in the bottom of the figure with the time markers.
The majority of primary capsules are routed to the class capsule corresponding to the correct symbol.
Among the two DR cases, the heat maps of the iteration 3 versions have a slightly higher contrast than that of the iteration 2 versions as shown in Fig.~\ref{fig:ccm}.
In other words, as the number of iterations increases, it seems that the distributions become less uniform.
The iteration 2 and 3 version of SDR models display the same phenomenon as the DR version.
However, the SDR model with iteration 1 seems to have different behaviors besides that $c$ is uniform at $t=1$.
The model routes the capsules to ``pau" more than any other version at $t=103$.
Moreover, at $t=137$, i.e., the end of the sentence, the model seems to route the majority of capsules to the class capsules corresponding to ``pau" rather than the correct symbol ``h\#" as explained earlier in this section.

\begin{figure}[ht!]
\centering
\subfloat[][$t$: 1 $\to$ 10]{
\includegraphics[width=1.0\linewidth]{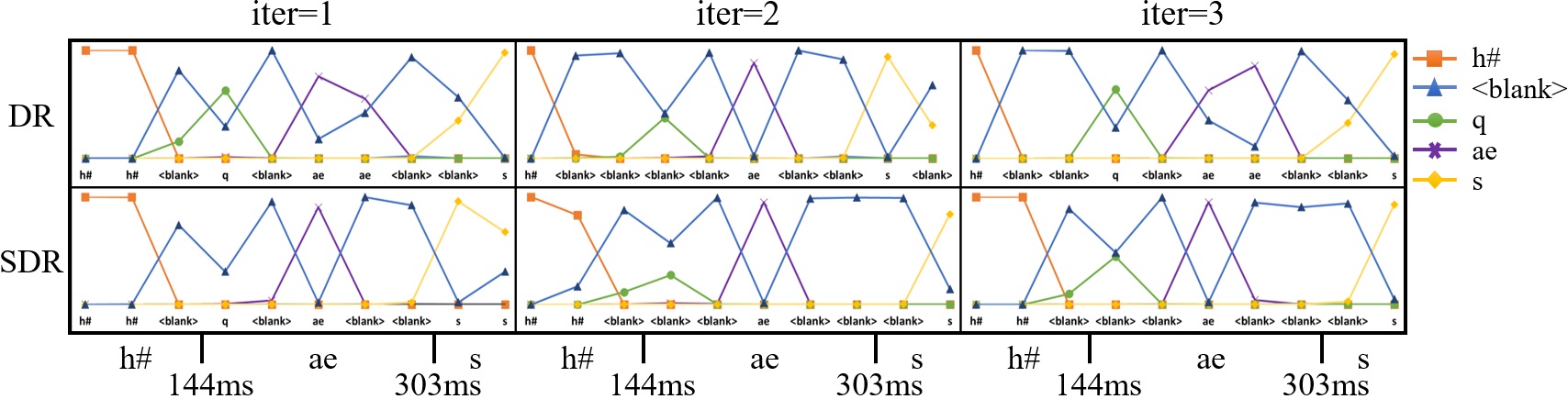}
\label{fig:softmax_start}
}
\qquad
\subfloat[][$t$: 127 $\to$ 137]{
\includegraphics[width=1.0\linewidth]{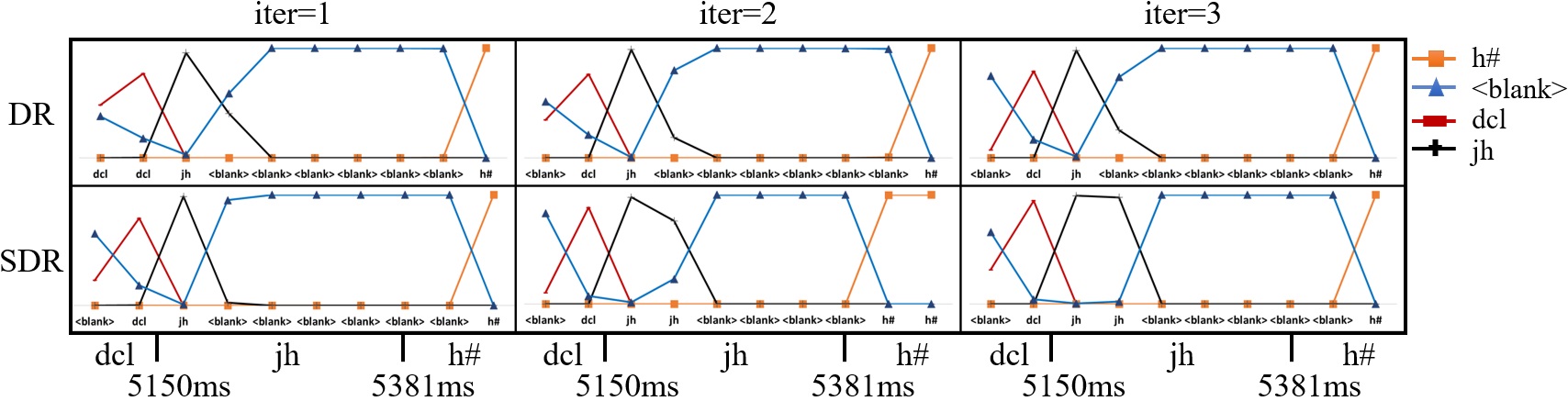}
\label{fig:softmax_end}
}
\caption{Softmax vectors depending on routing methods (horizontal-axis: time (ms), vertical-axis: probability)}
\label{fig:softmax_output}
\end{figure}

In order to see how the phenomenon affects alignment, we checked the 10 probability vectors each at the beginning and end of the sentence as shown in Fig~\ref{fig:softmax_output}.
Symbols are represented with different line shapes described in the right side of the figure.
The labels on the X-axis of each graph indicate a symbol with the highest probability, i.e., the phoneme sequence of the greedy decoding.
Reference time markers and symbols are written at the bottom of Fig~\ref{fig:softmax_start} and Fig~\ref{fig:softmax_end}.
For all the cases, the misalignment phenomenon of the iteration 1 version of SDR model is not observed.
It is not a big difference, but rather, the SDR model recognizes the symbol ``s" as fast as the iteration 2 version DR as shown in Fig~\ref{fig:softmax_start}.
The EOS symbol ``h\#" is also precisely recognized in all the cases as in Fig~\ref{fig:softmax_end}.
Even in the case of SDR when iteration is set to 1 at $t=137$, unlike the heat map where most of the primary capsules are routed to the class capsule to the symbol ``pau", the probability of ``h\#" is the largest.

\begin{table}[b!]
\centering
\begin{tabular}{cccc}
Routing & \multirow{2}{*}{Iteration} & \multicolumn{2}{c}{Edit distance rate (\%)} \\
Method  &                            & Valid & Test                \\ \hline
\multirow{2}{*}{DR}  & 2 & 0.14 & 0.14 \\
                     & 3 & 0.32 & 0.32 \\ \hline
\multirow{3}{*}{SDR} & 1 & 0.80 & 0.81 \\
                     & 2 & 0.56 & 0.57 \\
                     & 3 & 0.50 & 0.51
\end{tabular}
\caption{Edit distance rates without alignments between phoneme index sequences corresponding to the largest sum of coupling coefficients $c$ and the largest values of softmax vectors for the TIMIT corpus~\citep{timit}.}
\label{Tab:distance}
\end{table}

As in Table~\ref{Tab:distance}, we measured the edit distance rates without alignments by $\frac{\textrm{the number of substitutions}}{\textrm{the number of frames}}$ between phoneme index sequences corresponding to the largest sum of $c$ and the largest values of softmax vectors. As the number of iterations increases, the edit distance rates of DR models increases while the edit distance rates of SDR models decreases.

\section{Discussion}
The SDR method shows better PERs over the DR method on the TIMIT corpus.
Our intuition is that the delayed initialization, which initializes the current routing coefficients from the previous routing results, can act as a kind of regularization.
This is because the target labels corresponding to the consecutive time slices would be similar but not the same thus this initialization method can be regarded as adding noise to not overfit with the huge number of routing iterations.
Moreover, the misalignment phenomenon observed from the coupling coefficient maps of the SDR method does not necessarily result in errors of the decoding output.
We also observed that the differences in edit distances between the coupling coefficients and the softmax vectors, depending on the settings of the iteration numbers, does not affect the PERs for Valid and Test of TIMIT.
The length of the output capsules is determined by multiplying the prediction vectors and the routing coefficients thus the transformation matrices can have chances to learn how to recover the decoding errors caused by the misalignment.
The iteration 1 version of DR in Section 5 is an example where a correct phoneme sequence can be recognized with uniform routing coefficients, i.e., this is an example of correctly recognizing only with the transformation matrix.

The receptive field of the SRF models seem to be an important configuration to reduce the recognition error rates.
In addition, as label contexts get longer from phonemes to characters, longer receptive fields are required.
For faster training speed, we set capsulation blocks to stride 4 frames, thus receptive fields increase by $(\omega - 1) \times 4$ as each layer is stacked.
Furthermore, we set $M_H$ not to exceed 30 since even with the same number of parameters, when matrix multiplications between capsules increases, the learning speed becomes slower.
Even with these constraints, our current training system takes 14-hours to train an epoch for a corpus lower than a hundred hours on a  graphics processing unit (GPU) to show competitive word recognition accuracy. This is a very long training time when considering that ASR training systems based on a traditional neural network take 65.6-hours to train an epoch for a 10K-hour corpus with four GPUs~\citep{DBLP:conf/asru/KimSSHGKKKKLHGK19}. Regarding the decoding time, we observed only a minimal relationship between the input speech length and the decoding time of the proposed models. We believe that the proposed models can be trained and decoded faster by developing efficient implementations for calculating the prediction vectors which require matrix multiplications between every pair of capsules in the lower- and higher-levels per frame.

\section{Related work}
In this paper, we explored the potential capabilities of CapsNets for sequence encoding.
There have been many research related to CapsNets such as improving their routing mechanisms or applying them to various kinds of tasks.
In this section, we explain those attempts.
The concept of routing between capsules was first introduced to recognize pose information~\citep{DBLP:conf/icann/HintonKW11} and it was implemented in an auto encoder manner~\citep{DBLP:conf/nips/HintonZ93}.
DR~\citep{DBLP:conf/nips/SabourFH17} is the earliest attempt to apply capsules for image classification problems. It showed better accuracy, not only in the original MNIST~\citep{lecun-mnisthandwrittendigit-2010} dataset compared to CNNs, but also in highly overlapping digit cases in the MultiMNIST dataset.
The accuracy on the overlapping cases was on par with that of sequential attention models.
EM routing~\citep{DBLP:conf/iclr/HintonSF18}, which trains GMMs to cluster capsules, is a follow-up study of DR.
It not only releases the length constraint of the instantiation vectors by defining activations as separate scalars but also reduces the size of transformation matrices to the square root size by representing the instantiation parameters as matrices.
Despite its structural and theoretical advances, the EM routing method has shown noncompetitive accuracy and computational complexity compared to DR~\citep{Malmgren1314210, DBLP:conf/nips/HahnPK19}.
This is the reason why we did not adopt it in this study, but it still has a suitable structure to be applied to the proposed method.

In order to improve implementations of the pioneer studies, various modifications were proposed.
When it comes to iterative routing methods, since increasing the number of iterations can lead to unbalanced activations, ~\citep{DBLP:conf/iclr/Wang018} proposed an optimized DR method which applies an entropy regularizer to constrain the routing coefficient to be close to the uniform distribution.
A min-max normalization~\citep{DBLP:journals/corr/abs-1903-09662} was applied to resolve the performance degradation in DR caused by iterative usage of a softmax function as a normalization function for routing coefficients.
These two modifications are expected to help the SRF models learn more stably when the routing iteration numbers are increased by the long inputs.
For faster training, routing coefficients were initialized from the learnable weights~\citep{DBLP:conf/eccv/RamasingheAK18} and attempts to introduce the computationally efficient EM algorithm, which does not require calculating the variances, to DR~\citep{DBLP:series/sci/Zhang0W19} was studied.

Recently, CapsNet has been applied to relatively large datasets such as Canadian Institute For Advanced Research (CiFAR) 100~\citep{Krizhevsky09learningmultiple} by parallelizing iterative routing methods~\citep{DBLP:conf/iclr/TsaiSGS20} and showed competitive performance compared to ResNet~\citep{DBLP:conf/cvpr/HeZRS16}.
This algorithm can lessen the layer-wise computational dependencies thus we expect that it can be utilized to build up deeper CapsNet architectures while minimizing training and decoding time increases.
There is also self-routing~\citep{DBLP:conf/nips/HahnPK19}, which is a non-iterative routing method, that introduces the mixture-of-expect mechanism~\citep{DBLP:journals/neco/JacobsJNH91} to the routing method.
The method trains models solely depending on gradient-based weight updates thus it is expected to have less computational burden compared to other CapsNets.
By recurrently sharing the routing results, we expect to apply this algorithm to SRFs while maintaining the capability to learn sequential dependencies.

In addition to research which improves CapsNets themselves, there are various attempts to merge CapsNets or the routing mechanism with other models.
Especially for sequence inputs, CapsNets are combined with existing sequential models either as a successor block at the top of the LSTM layers~\citep{zhang-etal-2018-attention, DBLP:conf/icann/HePLHZ19}, the Transformer~\citep{DBLP:conf/nips/VaswaniSPUJGKP17} encoder blocks~\citep{liu-etal-2019-transformer}, and bidirectional encoder representations from Transformers (BERT)
~\citep{DBLP:conf/naacl/DevlinCLT19, DBLP:journals/jbi/SunYWZLW20} or as an intermediate block in between encoders and decoders which are made of LSTMs~\citep{wang-2019-towards}.
There are also attempts to put routing algorithms into attention methods and vice versa.
Routing mechanisms are adopted into self-attention based models to cluster similar information from the multi-head attentions~\citep{DBLP:conf/nlpcc/GuF19}.
In contrast, STAR-Caps~\citep{DBLP:conf/nips/AhmedT19} merges attention methods into the routing mechanism.

CapsNets have been actively applied to a variety of fields because of their outstanding image encoding abilities.
Thus, they are suitable for visual tracking~\citep{DBLP:journals/corr/abs-1902-10054}, object segmentation of medical images~\citep{DBLP:journals/corr/abs-1804-04241} and self-driving \citep{DBLP:journals/tjs/KimC19}.
CapsNets also can be easily applied to non-visual tasks such as knowledge graph embedding and link prediction because of their representations of conceptual hierarchy relationships~\citep{DBLP:conf/naacl/NguyenVNNP19, DBLP:conf/iclr/XinyiC19}.
For linguistic data, CapsNets were applied to text classification with k-means routing~\citep{DBLP:journals/corr/abs-1810-09177}, machine translation in an encoder-decoder manner~\citep{wang-2019-towards}, user intent detection~\citep{DBLP:conf/emnlp/XiaZYCY18, DBLP:conf/acl/ZhangLDFY19} and emotion detection using micro blogs~\citep{DBLP:journals/nca/ZhongLLCLDWZ20}.
Classification tasks using audio and speech data~\citep{DBLP:journals/corr/abs-1902-05069} have also been actively researched to detect sound events~\citep{DBLP:conf/eusipco/Iqbal0KW18, DBLP:journals/jstsp/VesperiniGPS19} and classify emotions~\citep{DBLP:conf/icassp/WuLCLYDMHWLM19}.
Electrocardiogram signal categorization~\citep{Jayasekara2019TimeCapsCT} is another interesting task where CapsNets can be applied to classify input sequential data.
Last but not least, isolated word recognition has been researched~\citep{DBLP:conf/interspeech/BaeK18, xiongyan2018master}.

\section{Conclusion}
We propose SRF, which is a novel framework to adapt CapsNets for encoding sequence data.
We believe, this is the first capsule-only structure for seq2seq recognition.
In the framework, input sequences are capsulized and sliced by the given window size.
Routing from lower to higher levels is performed for each slice by sharing two kinds of parameters over a whole sequence.
From the perspective of gradient-based optimization, the amount of required memory size can be controlled regardless of the length of input sequences by sharing the transformation matrix.
Moreover, by initializing routing coefficients based on the routing output of the previous slices, we could minimize additional computational burden caused by the routing iteration since the routing mechanism can be operated in a non-iterative manner for each slice.
The proposed method achieved competitive performance on the TIMIT and WSJ corpora in two aspects that are accuracy and streaming capabilities.
An area of improvement for the future is to research a new routing mechanism to reduce the algorithmic delays by enhancing the accuracy of unbalanced window configurations. 
In addition, it will be worth to study the possibility of a fully end-to-end ASR system, which can represent linguistic context, based on CapsNet-only architectures.

\bibliography{mybibfile}

\begin{thebibliography}{64}
\expandafter\ifx\csname natexlab\endcsname\relax\def\natexlab#1{#1}\fi
\providecommand{\url}[1]{\texttt{#1}}
\providecommand{\href}[2]{#2}
\providecommand{\path}[1]{#1}
\providecommand{\DOIprefix}{doi:}
\providecommand{\ArXivprefix}{arXiv:}
\providecommand{\URLprefix}{URL: }
\providecommand{\Pubmedprefix}{pmid:}
\providecommand{\doi}[1]{\href{http://dx.doi.org/#1}{\path{#1}}}
\providecommand{\Pubmed}[1]{\href{pmid:#1}{\path{#1}}}
\providecommand{\bibinfo}[2]{#2}
\ifx\xfnm\undefined \def\xfnm[#1]{\unskip,\space#1}\fi
\bibitem[{Abadi et~al.(2016)Abadi, Agarwal, Barham, Brevdo, Chen, Citro,
  Corrado, Davis, Dean, Devin, Ghemawat, Goodfellow, Harp, Irving, Isard, Jia,
  J{\'{o}}zefowicz, Kaiser, Kudlur, Levenberg, Man{\'{e}}, Monga, Moore,
  Murray, Olah, Schuster, Shlens, Steiner, Sutskever, Talwar, Tucker,
  Vanhoucke, Vasudevan, Vi{\'{e}}gas, Vinyals, Warden, Wattenberg, Wicke, Yu
  and Zheng}]{DBLP:journals/corr/AbadiABBCCCDDDG16}
\bibinfo{author}{Abadi\xfnm[ M.]}, \bibinfo{author}{Agarwal\xfnm[ A.]},
  \bibinfo{author}{Barham\xfnm[ P.]}, \bibinfo{author}{Brevdo\xfnm[ E.]},
  \bibinfo{author}{Chen\xfnm[ Z.]}, \bibinfo{author}{Citro\xfnm[ C.]},
  \bibinfo{author}{Corrado\xfnm[ G.S.]}, \bibinfo{author}{Davis\xfnm[ A.]},
  \bibinfo{author}{Dean\xfnm[ J.]}, \bibinfo{author}{Devin\xfnm[ M.]},
  \bibinfo{author}{Ghemawat\xfnm[ S.]}, \bibinfo{author}{Goodfellow\xfnm[
  I.J.]}, \bibinfo{author}{Harp\xfnm[ A.]}, \bibinfo{author}{Irving\xfnm[ G.]},
  \bibinfo{author}{Isard\xfnm[ M.]}, \bibinfo{author}{Jia\xfnm[ Y.]},
  \bibinfo{author}{J{\'{o}}zefowicz\xfnm[ R.]}, \bibinfo{author}{Kaiser\xfnm[
  L.]}, \bibinfo{author}{Kudlur\xfnm[ M.]}, \bibinfo{author}{Levenberg\xfnm[
  J.]}, \bibinfo{author}{Man{\'{e}}\xfnm[ D.]}, \bibinfo{author}{Monga\xfnm[
  R.]}, \bibinfo{author}{Moore\xfnm[ S.]}, \bibinfo{author}{Murray\xfnm[
  D.G.]}, \bibinfo{author}{Olah\xfnm[ C.]}, \bibinfo{author}{Schuster\xfnm[
  M.]}, \bibinfo{author}{Shlens\xfnm[ J.]}, \bibinfo{author}{Steiner\xfnm[
  B.]}, \bibinfo{author}{Sutskever\xfnm[ I.]}, \bibinfo{author}{Talwar\xfnm[
  K.]}, \bibinfo{author}{Tucker\xfnm[ P.A.]}, \bibinfo{author}{Vanhoucke\xfnm[
  V.]}, \bibinfo{author}{Vasudevan\xfnm[ V.]},
  \bibinfo{author}{Vi{\'{e}}gas\xfnm[ F.B.]}, \bibinfo{author}{Vinyals\xfnm[
  O.]}, \bibinfo{author}{Warden\xfnm[ P.]}, \bibinfo{author}{Wattenberg\xfnm[
  M.]}, \bibinfo{author}{Wicke\xfnm[ M.]}, \bibinfo{author}{Yu\xfnm[ Y.]},
  \bibinfo{author}{Zheng\xfnm[ X.]}.
\newblock \bibinfo{title}{Tensorflow: Large-scale machine learning on
  heterogeneous distributed systems}.
\newblock \bibinfo{journal}{CoRR}
  \bibinfo{year}{2016};\bibinfo{volume}{abs/1603.04467}.
\newblock \URLprefix \url{http://arxiv.org/abs/1603.04467}.
  \href{http://arxiv.org/abs/1603.04467}{\tt arXiv:1603.04467}.
\bibitem[{Ahmed and Torresani(2019)}]{DBLP:conf/nips/AhmedT19}
\bibinfo{author}{Ahmed\xfnm[ K.]}, \bibinfo{author}{Torresani\xfnm[ L.]}.
\newblock \bibinfo{title}{Star-caps: Capsule networks with straight-through
  attentive routing}.
\newblock In: \bibinfo{editor}{Wallach\xfnm[ H.M.]},
  \bibinfo{editor}{Larochelle\xfnm[ H.]}, \bibinfo{editor}{Beygelzimer\xfnm[
  A.]}, \bibinfo{editor}{d'Alch{\'{e}}{-}Buc\xfnm[ F.]},
  \bibinfo{editor}{Fox\xfnm[ E.B.]}, \bibinfo{editor}{Garnett\xfnm[ R.]},
  editors. \bibinfo{booktitle}{Advances in Neural Information Processing
  Systems 32: Annual Conference on Neural Information Processing Systems 2019,
  NeurIPS 2019, 8-14 December 2019, Vancouver, BC, Canada}.
  \bibinfo{year}{2019}. p. \bibinfo{pages}{9098--9107}.
\newblock \URLprefix
  \url{http://papers.nips.cc/paper/9110-star-caps-capsule-networks-with-straight-through-attentive-routing}.
\bibitem[{Ba et~al.(2016)Ba, Kiros and Hinton}]{DBLP:journals/corr/BaKH16}
\bibinfo{author}{Ba\xfnm[ L.J.]}, \bibinfo{author}{Kiros\xfnm[ J.R.]},
  \bibinfo{author}{Hinton\xfnm[ G.E.]}.
\newblock \bibinfo{title}{Layer normalization}.
\newblock \bibinfo{journal}{CoRR}
  \bibinfo{year}{2016};\bibinfo{volume}{abs/1607.06450}.
\newblock \URLprefix \url{http://arxiv.org/abs/1607.06450}.
  \href{http://arxiv.org/abs/1607.06450}{\tt arXiv:1607.06450}.
\bibitem[{Bae and Kim(2018)}]{DBLP:conf/interspeech/BaeK18}
\bibinfo{author}{Bae\xfnm[ J.]}, \bibinfo{author}{Kim\xfnm[ D.]}.
\newblock \bibinfo{title}{End-to-end speech command recognition with capsule
  network}.
\newblock In: \bibinfo{editor}{Yegnanarayana\xfnm[ B.]}, editor.
  \bibinfo{booktitle}{Interspeech 2018, 19th Annual Conference of the
  International Speech Communication Association, Hyderabad, India, 2-6
  September 2018}. \bibinfo{publisher}{{ISCA}}; \bibinfo{year}{2018}. p.
  \bibinfo{pages}{776--780}.
\newblock \URLprefix \url{https://doi.org/10.21437/Interspeech.2018-1888}.
  \DOIprefix\doi{10.21437/Interspeech.2018-1888}.
\bibitem[{Consortium and Group(1994)}]{wsj1}
\bibinfo{author}{Consortium\xfnm[ L.D.]}, \bibinfo{author}{Group\xfnm[
  N.M.I.]}.
\newblock \bibinfo{title}{Csr-ii (wsj1) complete}.
\newblock \bibinfo{year}{1994}.
\newblock \URLprefix \url{https://catalog.ldc.upenn.edu/LDC94S13A}.
\bibitem[{Devlin et~al.(2019)Devlin, Chang, Lee and
  Toutanova}]{DBLP:conf/naacl/DevlinCLT19}
\bibinfo{author}{Devlin\xfnm[ J.]}, \bibinfo{author}{Chang\xfnm[ M.]},
  \bibinfo{author}{Lee\xfnm[ K.]}, \bibinfo{author}{Toutanova\xfnm[ K.]}.
\newblock \bibinfo{title}{{BERT:} pre-training of deep bidirectional
  transformers for language understanding}.
\newblock In: \bibinfo{editor}{Burstein\xfnm[ J.]},
  \bibinfo{editor}{Doran\xfnm[ C.]}, \bibinfo{editor}{Solorio\xfnm[ T.]},
  editors. \bibinfo{booktitle}{Proceedings of the 2019 Conference of the North
  American Chapter of the Association for Computational Linguistics: Human
  Language Technologies, {NAACL-HLT} 2019, Minneapolis, MN, USA, June 2-7,
  2019, Volume 1 (Long and Short Papers)}. \bibinfo{publisher}{Association for
  Computational Linguistics}; \bibinfo{year}{2019}. p.
  \bibinfo{pages}{4171--4186}.
\newblock \URLprefix \url{https://doi.org/10.18653/v1/n19-1423}.
  \DOIprefix\doi{10.18653/v1/n19-1423}.
\bibitem[{{Dong} et~al.(2018){Dong}, {Xu} and {Xu}}]{8462506}
\bibinfo{author}{{Dong}\xfnm[ L.]}, \bibinfo{author}{{Xu}\xfnm[ S.]},
  \bibinfo{author}{{Xu}\xfnm[ B.]}.
\newblock \bibinfo{title}{Speech-transformer: A no-recurrence
  sequence-to-sequence model for speech recognition}.
\newblock In: \bibinfo{booktitle}{2018 IEEE International Conference on
  Acoustics, Speech and Signal Processing (ICASSP)}. \bibinfo{year}{2018}. p.
  \bibinfo{pages}{5884--5888}.
\bibitem[{Garofolo et~al.(1993{\natexlab{a}})Garofolo, Graff, Paul and
  Pallett}]{wsj0}
\bibinfo{author}{Garofolo\xfnm[ J.S.]}, \bibinfo{author}{Graff\xfnm[ D.]},
  \bibinfo{author}{Paul\xfnm[ D.]}, \bibinfo{author}{Pallett\xfnm[ D.]}.
\newblock \bibinfo{title}{Csr-i (wsj0) complete}.
\newblock \bibinfo{year}{1993}{\natexlab{a}}.
\newblock \URLprefix \url{https://catalog.ldc.upenn.edu/LDC93S6A}.
\bibitem[{Garofolo et~al.(1993{\natexlab{b}})Garofolo, Lamel, Fisher, Fiscus,
  Pallett, Dahlgren and Zue}]{timit}
\bibinfo{author}{Garofolo\xfnm[ J.S.]}, \bibinfo{author}{Lamel\xfnm[ L.F.]},
  \bibinfo{author}{Fisher\xfnm[ W.M.]}, \bibinfo{author}{Fiscus\xfnm[ J.G.]},
  \bibinfo{author}{Pallett\xfnm[ D.S.]}, \bibinfo{author}{Dahlgren\xfnm[
  N.L.]}, \bibinfo{author}{Zue\xfnm[ V.]}.
\newblock \bibinfo{title}{Darpa timit acoustic phonetic continuous speech
  corpus cdrom}.
\newblock \bibinfo{year}{1993}{\natexlab{b}}.
\bibitem[{Glorot and Bengio(2010)}]{DBLP:journals/jmlr/GlorotB10}
\bibinfo{author}{Glorot\xfnm[ X.]}, \bibinfo{author}{Bengio\xfnm[ Y.]}.
\newblock \bibinfo{title}{Understanding the difficulty of training deep
  feedforward neural networks}.
\newblock In: \bibinfo{editor}{Teh\xfnm[ Y.W.]},
  \bibinfo{editor}{Titterington\xfnm[ D.M.]}, editors.
  \bibinfo{booktitle}{Proceedings of the Thirteenth International Conference on
  Artificial Intelligence and Statistics, {AISTATS} 2010, Chia Laguna Resort,
  Sardinia, Italy, May 13-15, 2010}. \bibinfo{publisher}{JMLR.org};
  volume~\bibinfo{volume}{9} of \textit{\bibinfo{series}{{JMLR} Proceedings}};
  \bibinfo{year}{2010}. p. \bibinfo{pages}{249--256}.
\newblock \URLprefix \url{http://proceedings.mlr.press/v9/glorot10a.html}.
\bibitem[{Goodfellow et~al.(2013)Goodfellow, Warde{-}Farley, Mirza, Courville
  and Bengio}]{DBLP:journals/corr/abs-1302-4389}
\bibinfo{author}{Goodfellow\xfnm[ I.J.]}, \bibinfo{author}{Warde{-}Farley\xfnm[
  D.]}, \bibinfo{author}{Mirza\xfnm[ M.]}, \bibinfo{author}{Courville\xfnm[
  A.C.]}, \bibinfo{author}{Bengio\xfnm[ Y.]}.
\newblock \bibinfo{title}{Maxout networks}.
\newblock \bibinfo{journal}{CoRR}
  \bibinfo{year}{2013};\bibinfo{volume}{abs/1302.4389}.
\newblock \URLprefix \url{http://arxiv.org/abs/1302.4389}.
  \href{http://arxiv.org/abs/1302.4389}{\tt arXiv:1302.4389}.
\bibitem[{Graves et~al.(2006)Graves, Fern{\'{a}}ndez, Gomez and
  Schmidhuber}]{DBLP:conf/icml/GravesFGS06}
\bibinfo{author}{Graves\xfnm[ A.]}, \bibinfo{author}{Fern{\'{a}}ndez\xfnm[
  S.]}, \bibinfo{author}{Gomez\xfnm[ F.J.]}, \bibinfo{author}{Schmidhuber\xfnm[
  J.]}.
\newblock \bibinfo{title}{Connectionist temporal ification: labelling
  unsegmented sequence data with recurrent neural networks}.
\newblock In: \bibinfo{editor}{Cohen\xfnm[ W.W.]}, \bibinfo{editor}{Moore\xfnm[
  A.W.]}, editors. \bibinfo{booktitle}{Machine Learning, Proceedings of the
  Twenty-Third International Conference {(ICML} 2006), Pittsburgh,
  Pennsylvania, USA, June 25-29, 2006}. \bibinfo{publisher}{{ACM}}; volume
  \bibinfo{volume}{148} of \textit{\bibinfo{series}{{ACM} International
  Conference Proceeding Series}}; \bibinfo{year}{2006}. p.
  \bibinfo{pages}{369--376}.
\newblock \URLprefix \url{https://doi.org/10.1145/1143844.1143891}.
  \DOIprefix\doi{10.1145/1143844.1143891}.
\bibitem[{Graves et~al.(2013)Graves, Mohamed and
  Hinton}]{DBLP:conf/icassp/GravesMH13}
\bibinfo{author}{Graves\xfnm[ A.]}, \bibinfo{author}{Mohamed\xfnm[ A.]},
  \bibinfo{author}{Hinton\xfnm[ G.E.]}.
\newblock \bibinfo{title}{Speech recognition with deep recurrent neural
  networks}.
\newblock In: \bibinfo{booktitle}{{IEEE} International Conference on Acoustics,
  Speech and Signal Processing, {ICASSP} 2013, Vancouver, BC, Canada, May
  26-31, 2013}. \bibinfo{publisher}{{IEEE}}; \bibinfo{year}{2013}. p.
  \bibinfo{pages}{6645--6649}.
\newblock \URLprefix \url{https://doi.org/10.1109/ICASSP.2013.6638947}.
  \DOIprefix\doi{10.1109/ICASSP.2013.6638947}.
\bibitem[{Gu and Feng(2019)}]{DBLP:conf/nlpcc/GuF19}
\bibinfo{author}{Gu\xfnm[ S.]}, \bibinfo{author}{Feng\xfnm[ Y.]}.
\newblock \bibinfo{title}{Improving multi-head attention with capsule
  networks}.
\newblock In: \bibinfo{editor}{Tang\xfnm[ J.]}, \bibinfo{editor}{Kan\xfnm[
  M.]}, \bibinfo{editor}{Zhao\xfnm[ D.]}, \bibinfo{editor}{Li\xfnm[ S.]},
  \bibinfo{editor}{Zan\xfnm[ H.]}, editors. \bibinfo{booktitle}{Natural
  Language Processing and Chinese Computing - 8th {CCF} International
  Conference, {NLPCC} 2019, Dunhuang, China, October 9-14, 2019, Proceedings,
  Part {I}}. \bibinfo{publisher}{Springer}; volume \bibinfo{volume}{11838} of
  \textit{\bibinfo{series}{Lecture Notes in Computer Science}};
  \bibinfo{year}{2019}. p. \bibinfo{pages}{314--326}.
\newblock \URLprefix \url{https://doi.org/10.1007/978-3-030-32233-5\_25}.
  \DOIprefix\doi{10.1007/978-3-030-32233-5\_25}.
\bibitem[{Hahn et~al.(2019)Hahn, Pyeon and Kim}]{DBLP:conf/nips/HahnPK19}
\bibinfo{author}{Hahn\xfnm[ T.]}, \bibinfo{author}{Pyeon\xfnm[ M.]},
  \bibinfo{author}{Kim\xfnm[ G.]}.
\newblock \bibinfo{title}{Self-routing capsule networks}.
\newblock In: \bibinfo{editor}{Wallach\xfnm[ H.M.]},
  \bibinfo{editor}{Larochelle\xfnm[ H.]}, \bibinfo{editor}{Beygelzimer\xfnm[
  A.]}, \bibinfo{editor}{d'Alch{\'{e}}{-}Buc\xfnm[ F.]},
  \bibinfo{editor}{Fox\xfnm[ E.B.]}, \bibinfo{editor}{Garnett\xfnm[ R.]},
  editors. \bibinfo{booktitle}{Advances in Neural Information Processing
  Systems 32: Annual Conference on Neural Information Processing Systems 2019,
  NeurIPS 2019, 8-14 December 2019, Vancouver, BC, Canada}.
  \bibinfo{year}{2019}. p. \bibinfo{pages}{7656--7665}.
\newblock \URLprefix
  \url{http://papers.nips.cc/paper/8982-self-routing-capsule-networks}.
\bibitem[{He et~al.(2019)He, Peng, Le, He and Zhu}]{DBLP:conf/icann/HePLHZ19}
\bibinfo{author}{He\xfnm[ C.]}, \bibinfo{author}{Peng\xfnm[ L.]},
  \bibinfo{author}{Le\xfnm[ Y.]}, \bibinfo{author}{He\xfnm[ J.]},
  \bibinfo{author}{Zhu\xfnm[ X.]}.
\newblock \bibinfo{title}{Secaps: {A} sequence enhanced capsule model for
  charge prediction}.
\newblock In: \bibinfo{editor}{Tetko\xfnm[ I.V.]},
  \bibinfo{editor}{Kurkov{\'{a}}\xfnm[ V.]}, \bibinfo{editor}{Karpov\xfnm[
  P.]}, \bibinfo{editor}{Theis\xfnm[ F.J.]}, editors.
  \bibinfo{booktitle}{Artificial Neural Networks and Machine Learning - {ICANN}
  2019: Text and Time Series - 28th International Conference on Artificial
  Neural Networks, Munich, Germany, September 17-19, 2019, Proceedings, Part
  {IV}}. \bibinfo{publisher}{Springer}; volume \bibinfo{volume}{11730} of
  \textit{\bibinfo{series}{Lecture Notes in Computer Science}};
  \bibinfo{year}{2019}. p. \bibinfo{pages}{227--239}.
\newblock \URLprefix \url{https://doi.org/10.1007/978-3-030-30490-4\_19}.
  \DOIprefix\doi{10.1007/978-3-030-30490-4\_19}.
\bibitem[{He et~al.(2016)He, Zhang, Ren and Sun}]{DBLP:conf/cvpr/HeZRS16}
\bibinfo{author}{He\xfnm[ K.]}, \bibinfo{author}{Zhang\xfnm[ X.]},
  \bibinfo{author}{Ren\xfnm[ S.]}, \bibinfo{author}{Sun\xfnm[ J.]}.
\newblock \bibinfo{title}{Deep residual learning for image recognition}.
\newblock In: \bibinfo{booktitle}{2016 {IEEE} Conference on Computer Vision and
  Pattern Recognition, {CVPR} 2016, Las Vegas, NV, USA, June 27-30, 2016}.
  \bibinfo{publisher}{{IEEE} Computer Society}; \bibinfo{year}{2016}. p.
  \bibinfo{pages}{770--778}.
\newblock \URLprefix \url{https://doi.org/10.1109/CVPR.2016.90}.
  \DOIprefix\doi{10.1109/CVPR.2016.90}.
\bibitem[{Hinton et~al.(2012)Hinton, Deng, Yu, Dahl, rahman Mohamed, Jaitly,
  Senior, Vanhoucke, Nguyen, Sainath and Kingsbury}]{38131}
\bibinfo{author}{Hinton\xfnm[ G.]}, \bibinfo{author}{Deng\xfnm[ L.]},
  \bibinfo{author}{Yu\xfnm[ D.]}, \bibinfo{author}{Dahl\xfnm[ G.]},
  \bibinfo{author}{rahman Mohamed\xfnm[ A.]}, \bibinfo{author}{Jaitly\xfnm[
  N.]}, \bibinfo{author}{Senior\xfnm[ A.]}, \bibinfo{author}{Vanhoucke\xfnm[
  V.]}, \bibinfo{author}{Nguyen\xfnm[ P.]}, \bibinfo{author}{Sainath\xfnm[
  T.]}, \bibinfo{author}{Kingsbury\xfnm[ B.]}.
\newblock \bibinfo{title}{Deep neural networks for acoustic modeling in speech
  recognition}.
\newblock \bibinfo{journal}{Signal Processing Magazine} \bibinfo{year}{2012};.
\bibitem[{Hinton et~al.(2011)Hinton, Krizhevsky and
  Wang}]{DBLP:conf/icann/HintonKW11}
\bibinfo{author}{Hinton\xfnm[ G.E.]}, \bibinfo{author}{Krizhevsky\xfnm[ A.]},
  \bibinfo{author}{Wang\xfnm[ S.D.]}.
\newblock \bibinfo{title}{Transforming auto-encoders}.
\newblock In: \bibinfo{editor}{Honkela\xfnm[ T.]}, \bibinfo{editor}{Duch\xfnm[
  W.]}, \bibinfo{editor}{Girolami\xfnm[ M.A.]}, \bibinfo{editor}{Kaski\xfnm[
  S.]}, editors. \bibinfo{booktitle}{Artificial Neural Networks and Machine
  Learning - {ICANN} 2011 - 21st International Conference on Artificial Neural
  Networks, Espoo, Finland, June 14-17, 2011, Proceedings, Part {I}}.
  \bibinfo{publisher}{Springer}; volume \bibinfo{volume}{6791} of
  \textit{\bibinfo{series}{Lecture Notes in Computer Science}};
  \bibinfo{year}{2011}. p. \bibinfo{pages}{44--51}.
\newblock \URLprefix \url{https://doi.org/10.1007/978-3-642-21735-7\_6}.
  \DOIprefix\doi{10.1007/978-3-642-21735-7\_6}.
\bibitem[{Hinton et~al.(2018)Hinton, Sabour and
  Frosst}]{DBLP:conf/iclr/HintonSF18}
\bibinfo{author}{Hinton\xfnm[ G.E.]}, \bibinfo{author}{Sabour\xfnm[ S.]},
  \bibinfo{author}{Frosst\xfnm[ N.]}.
\newblock \bibinfo{title}{Matrix capsules with {EM} routing}.
\newblock In: \bibinfo{booktitle}{6th International Conference on Learning
  Representations, {ICLR} 2018, Vancouver, BC, Canada, April 30 - May 3, 2018,
  Conference Track Proceedings}. \bibinfo{publisher}{OpenReview.net};
  \bibinfo{year}{2018}. \URLprefix
  \url{https://openreview.net/forum?id=HJWLfGWRb}.
\bibitem[{Hinton and Zemel(1993)}]{DBLP:conf/nips/HintonZ93}
\bibinfo{author}{Hinton\xfnm[ G.E.]}, \bibinfo{author}{Zemel\xfnm[ R.S.]}.
\newblock \bibinfo{title}{Autoencoders, minimum description length and
  helmholtz free energy}.
\newblock In: \bibinfo{editor}{Cowan\xfnm[ J.D.]},
  \bibinfo{editor}{Tesauro\xfnm[ G.]}, \bibinfo{editor}{Alspector\xfnm[ J.]},
  editors. \bibinfo{booktitle}{Advances in Neural Information Processing
  Systems 6, [7th {NIPS} Conference, Denver, Colorado, USA, 1993]}.
  \bibinfo{publisher}{Morgan Kaufmann}; \bibinfo{year}{1993}. p.
  \bibinfo{pages}{3--10}.
\newblock \URLprefix
  \url{http://papers.nips.cc/paper/798-autoencoders-minimum-description-length-and-helmholtz-free-energy}.
\bibitem[{Hochreiter and Schmidhuber(1997)}]{DBLP:journals/neco/HochreiterS97}
\bibinfo{author}{Hochreiter\xfnm[ S.]}, \bibinfo{author}{Schmidhuber\xfnm[
  J.]}.
\newblock \bibinfo{title}{Long short-term memory}.
\newblock \bibinfo{journal}{Neural Computation}
  \bibinfo{year}{1997};\bibinfo{volume}{9}(\bibinfo{number}{8}):\bibinfo{pages}{1735--1780}.
\newblock \URLprefix \url{https://doi.org/10.1162/neco.1997.9.8.1735}.
  \DOIprefix\doi{10.1162/neco.1997.9.8.1735}.
\bibitem[{Ioffe and Szegedy(2015)}]{DBLP:conf/icml/IoffeS15}
\bibinfo{author}{Ioffe\xfnm[ S.]}, \bibinfo{author}{Szegedy\xfnm[ C.]}.
\newblock \bibinfo{title}{Batch normalization: Accelerating deep network
  training by reducing internal covariate shift}.
\newblock In: \bibinfo{editor}{Bach\xfnm[ F.R.]}, \bibinfo{editor}{Blei\xfnm[
  D.M.]}, editors. \bibinfo{booktitle}{Proceedings of the 32nd International
  Conference on Machine Learning, {ICML} 2015, Lille, France, 6-11 July 2015}.
  \bibinfo{publisher}{JMLR.org}; volume~\bibinfo{volume}{37} of
  \textit{\bibinfo{series}{{JMLR} Workshop and Conference Proceedings}};
  \bibinfo{year}{2015}. p. \bibinfo{pages}{448--456}.
\newblock \URLprefix \url{http://proceedings.mlr.press/v37/ioffe15.html}.
\bibitem[{Iqbal et~al.(2018)Iqbal, Xu, Kong and
  Wang}]{DBLP:conf/eusipco/Iqbal0KW18}
\bibinfo{author}{Iqbal\xfnm[ T.]}, \bibinfo{author}{Xu\xfnm[ Y.]},
  \bibinfo{author}{Kong\xfnm[ Q.]}, \bibinfo{author}{Wang\xfnm[ W.]}.
\newblock \bibinfo{title}{Capsule routing for sound event detection}.
\newblock In: \bibinfo{booktitle}{26th European Signal Processing Conference,
  {EUSIPCO} 2018, Roma, Italy, September 3-7, 2018}.
  \bibinfo{publisher}{{IEEE}}; \bibinfo{year}{2018}. p.
  \bibinfo{pages}{2255--2259}.
\newblock \URLprefix \url{https://doi.org/10.23919/EUSIPCO.2018.8553198}.
  \DOIprefix\doi{10.23919/EUSIPCO.2018.8553198}.
\bibitem[{Jacobs et~al.(1991)Jacobs, Jordan, Nowlan and
  Hinton}]{DBLP:journals/neco/JacobsJNH91}
\bibinfo{author}{Jacobs\xfnm[ R.A.]}, \bibinfo{author}{Jordan\xfnm[ M.I.]},
  \bibinfo{author}{Nowlan\xfnm[ S.J.]}, \bibinfo{author}{Hinton\xfnm[ G.E.]}.
\newblock \bibinfo{title}{Adaptive mixtures of local experts}.
\newblock \bibinfo{journal}{Neural Computation}
  \bibinfo{year}{1991};\bibinfo{volume}{3}(\bibinfo{number}{1}):\bibinfo{pages}{79--87}.
\newblock \URLprefix \url{https://doi.org/10.1162/neco.1991.3.1.79}.
  \DOIprefix\doi{10.1162/neco.1991.3.1.79}.
\bibitem[{Jain(2019)}]{DBLP:journals/corr/abs-1902-05069}
\bibinfo{author}{Jain\xfnm[ R.]}.
\newblock \bibinfo{title}{Improving performance and inference on audio
  classification tasks using capsule networks}.
\newblock \bibinfo{journal}{CoRR}
  \bibinfo{year}{2019};\bibinfo{volume}{abs/1902.05069}.
\newblock \URLprefix \url{http://arxiv.org/abs/1902.05069}.
  \href{http://arxiv.org/abs/1902.05069}{\tt arXiv:1902.05069}.
\bibitem[{Jayasekara et~al.(2019)Jayasekara, Jayasundara, Rajasegaran,
  Jayasekara, Seneviratne and Rodrigo}]{Jayasekara2019TimeCapsCT}
\bibinfo{author}{Jayasekara\xfnm[ H.]}, \bibinfo{author}{Jayasundara\xfnm[
  V.]}, \bibinfo{author}{Rajasegaran\xfnm[ J.]},
  \bibinfo{author}{Jayasekara\xfnm[ S.]}, \bibinfo{author}{Seneviratne\xfnm[
  S.]}, \bibinfo{author}{Rodrigo\xfnm[ R.]}.
\newblock \bibinfo{title}{Timecaps: Capturing time series data with capsule
  networks}.
\newblock \bibinfo{journal}{ArXiv}
  \bibinfo{year}{2019};\bibinfo{volume}{abs/1911.11800}.
\bibitem[{Kim et~al.(2019)Kim, Shin, Singh, Heck, Gowda, Kim, Kim, Kumar, Kim,
  Lee, Han, Garg and Kim}]{DBLP:conf/asru/KimSSHGKKKKLHGK19}
\bibinfo{author}{Kim\xfnm[ C.]}, \bibinfo{author}{Shin\xfnm[ M.]},
  \bibinfo{author}{Singh\xfnm[ S.]}, \bibinfo{author}{Heck\xfnm[ L.]},
  \bibinfo{author}{Gowda\xfnm[ D.]}, \bibinfo{author}{Kim\xfnm[ S.]},
  \bibinfo{author}{Kim\xfnm[ K.]}, \bibinfo{author}{Kumar\xfnm[ M.]},
  \bibinfo{author}{Kim\xfnm[ J.]}, \bibinfo{author}{Lee\xfnm[ K.]},
  \bibinfo{author}{Han\xfnm[ C.]}, \bibinfo{author}{Garg\xfnm[ A.]},
  \bibinfo{author}{Kim\xfnm[ E.]}.
\newblock \bibinfo{title}{End-to-end training of a large vocabulary end-to-end
  speech recognition system}.
\newblock In: \bibinfo{booktitle}{{IEEE} Automatic Speech Recognition and
  Understanding Workshop, {ASRU} 2019, Singapore, December 14-18, 2019}.
  \bibinfo{publisher}{{IEEE}}; \bibinfo{year}{2019}. p.
  \bibinfo{pages}{562--569}.
\newblock \URLprefix \url{https://doi.org/10.1109/ASRU46091.2019.9003976}.
  \DOIprefix\doi{10.1109/ASRU46091.2019.9003976}.
\bibitem[{Kim and Chi(2019)}]{DBLP:journals/tjs/KimC19}
\bibinfo{author}{Kim\xfnm[ M.]}, \bibinfo{author}{Chi\xfnm[ S.]}.
\newblock \bibinfo{title}{Detection of centerline crossing in abnormal driving
  using capsnet}.
\newblock \bibinfo{journal}{J Supercomput}
  \bibinfo{year}{2019};\bibinfo{volume}{75}(\bibinfo{number}{1}):\bibinfo{pages}{189--196}.
\newblock \URLprefix \url{https://doi.org/10.1007/s11227-018-2459-6}.
  \DOIprefix\doi{10.1007/s11227-018-2459-6}.
\bibitem[{Kingma and Ba(2015)}]{DBLP:journals/corr/KingmaB14}
\bibinfo{author}{Kingma\xfnm[ D.P.]}, \bibinfo{author}{Ba\xfnm[ J.]}.
\newblock \bibinfo{title}{Adam: {A} method for stochastic optimization}.
\newblock In: \bibinfo{editor}{Bengio\xfnm[ Y.]}, \bibinfo{editor}{LeCun\xfnm[
  Y.]}, editors. \bibinfo{booktitle}{3rd International Conference on Learning
  Representations, {ICLR} 2015, San Diego, CA, USA, May 7-9, 2015, Conference
  Track Proceedings}. \bibinfo{year}{2015}. \URLprefix
  \url{http://arxiv.org/abs/1412.6980}.
\bibitem[{Krizhevsky(2009)}]{Krizhevsky09learningmultiple}
\bibinfo{author}{Krizhevsky\xfnm[ A.]}.
\newblock \bibinfo{title}{Learning multiple layers of features from tiny
  images}.
\newblock \bibinfo{type}{Technical Report}; University of Toronto;
  \bibinfo{year}{2009}.
\bibitem[{LaLonde and Bagci(2018)}]{DBLP:journals/corr/abs-1804-04241}
\bibinfo{author}{LaLonde\xfnm[ R.]}, \bibinfo{author}{Bagci\xfnm[ U.]}.
\newblock \bibinfo{title}{Capsules for object segmentation}.
\newblock \bibinfo{journal}{CoRR}
  \bibinfo{year}{2018};\bibinfo{volume}{abs/1804.04241}.
\newblock \URLprefix \url{http://arxiv.org/abs/1804.04241}.
  \href{http://arxiv.org/abs/1804.04241}{\tt arXiv:1804.04241}.
\bibitem[{LeCun and Cortes(2010)}]{lecun-mnisthandwrittendigit-2010}
\bibinfo{author}{LeCun\xfnm[ Y.]}, \bibinfo{author}{Cortes\xfnm[ C.]}.
\newblock \bibinfo{title}{{MNIST} handwritten digit database}.
\newblock \bibinfo{journal}{empty} \bibinfo{year}{2010};\URLprefix
  \url{http://yann.lecun.com/exdb/mnist/}.
\bibitem[{Lee and Hon(1989)}]{DBLP:journals/tsp/LeeH89}
\bibinfo{author}{Lee\xfnm[ K.]}, \bibinfo{author}{Hon\xfnm[ H.]}.
\newblock \bibinfo{title}{Speaker-independent phone recognition using hidden
  markov models}.
\newblock \bibinfo{journal}{{IEEE} Trans Acoust Speech Signal Process}
  \bibinfo{year}{1989};\bibinfo{volume}{37}(\bibinfo{number}{11}):\bibinfo{pages}{1641--1648}.
\newblock \URLprefix \url{https://doi.org/10.1109/29.46546}.
  \DOIprefix\doi{10.1109/29.46546}.
\bibitem[{Lenssen et~al.(2018)Lenssen, Fey and Libuschewski}]{NIPS2018_8100}
\bibinfo{author}{Lenssen\xfnm[ J.E.]}, \bibinfo{author}{Fey\xfnm[ M.]},
  \bibinfo{author}{Libuschewski\xfnm[ P.]}.
\newblock \bibinfo{title}{Group equivariant capsule networks}.
\newblock In: \bibinfo{editor}{Bengio\xfnm[ S.]},
  \bibinfo{editor}{Wallach\xfnm[ H.]}, \bibinfo{editor}{Larochelle\xfnm[ H.]},
  \bibinfo{editor}{Grauman\xfnm[ K.]}, \bibinfo{editor}{Cesa-Bianchi\xfnm[
  N.]}, \bibinfo{editor}{Garnett\xfnm[ R.]}, editors.
  \bibinfo{booktitle}{Advances in Neural Information Processing Systems 31}.
  \bibinfo{publisher}{Curran Associates, Inc.}; \bibinfo{year}{2018}. p.
  \bibinfo{pages}{8844--8853}.
\newblock \URLprefix
  \url{http://papers.nips.cc/paper/8100-group-equivariant-capsule-networks.pdf}.
\bibitem[{Liu et~al.(2019)Liu, Lin, Liu, Xu, Ren, Diao and
  Yang}]{liu-etal-2019-transformer}
\bibinfo{author}{Liu\xfnm[ J.]}, \bibinfo{author}{Lin\xfnm[ H.]},
  \bibinfo{author}{Liu\xfnm[ X.]}, \bibinfo{author}{Xu\xfnm[ B.]},
  \bibinfo{author}{Ren\xfnm[ Y.]}, \bibinfo{author}{Diao\xfnm[ Y.]},
  \bibinfo{author}{Yang\xfnm[ L.]}.
\newblock \bibinfo{title}{Transformer-based capsule network for stock movement
  prediction}.
\newblock In: \bibinfo{booktitle}{Proceedings of the First Workshop on
  Financial Technology and Natural Language Processing}.
  \bibinfo{address}{Macao, China}; \bibinfo{year}{2019}. p.
  \bibinfo{pages}{66--73}.
\newblock \URLprefix \url{https://www.aclweb.org/anthology/W19-5511}.
\bibitem[{Ma and Wu(2019)}]{DBLP:journals/corr/abs-1902-10054}
\bibinfo{author}{Ma\xfnm[ D.]}, \bibinfo{author}{Wu\xfnm[ X.]}.
\newblock \bibinfo{title}{Tcdcaps: Visual tracking via cascaded dense
  capsules}.
\newblock \bibinfo{journal}{CoRR}
  \bibinfo{year}{2019};\bibinfo{volume}{abs/1902.10054}.
\newblock \URLprefix \url{http://arxiv.org/abs/1902.10054}.
  \href{http://arxiv.org/abs/1902.10054}{\tt arXiv:1902.10054};
  \bibinfo{note}{withdrawn.}
\bibitem[{Malmgren(2019)}]{Malmgren1314210}
\bibinfo{author}{Malmgren\xfnm[ C.]}.
\newblock \bibinfo{title}{A Comparative Study of Routing Methods in Capsule
  Networks}.
\newblock Master's thesis; Linköping University, Computer Vision;
  \bibinfo{year}{2019}.
\bibitem[{{Marchisio} et~al.(2020){Marchisio}, {Bussolino}, {Colucci}, {Hanif},
  {Martina}, {Masera} and {Shafique}}]{9207533}
\bibinfo{author}{{Marchisio}\xfnm[ A.]}, \bibinfo{author}{{Bussolino}\xfnm[
  B.]}, \bibinfo{author}{{Colucci}\xfnm[ A.]}, \bibinfo{author}{{Hanif}\xfnm[
  M.A.]}, \bibinfo{author}{{Martina}\xfnm[ M.]},
  \bibinfo{author}{{Masera}\xfnm[ G.]}, \bibinfo{author}{{Shafique}\xfnm[ M.]}.
\newblock \bibinfo{title}{Fastrcaps: An integrated framework for fast yet
  accurate training of capsule networks}.
\newblock In: \bibinfo{booktitle}{2020 International Joint Conference on Neural
  Networks (IJCNN)}. \bibinfo{year}{2020}. p. \bibinfo{pages}{1--8}.
\newblock \DOIprefix\doi{10.1109/IJCNN48605.2020.9207533}.
\bibitem[{{Moritz} et~al.(2020){Moritz}, {Hori} and {Le}}]{9054476}
\bibinfo{author}{{Moritz}\xfnm[ N.]}, \bibinfo{author}{{Hori}\xfnm[ T.]},
  \bibinfo{author}{{Le}\xfnm[ J.]}.
\newblock \bibinfo{title}{Streaming automatic speech recognition with the
  transformer model}.
\newblock In: \bibinfo{booktitle}{ICASSP 2020 - 2020 IEEE International
  Conference on Acoustics, Speech and Signal Processing (ICASSP)}.
  \bibinfo{year}{2020}. p. \bibinfo{pages}{6074--6078}.
\bibitem[{Nguyen et~al.(2019)Nguyen, Vu, Nguyen, Nguyen and
  Phung}]{DBLP:conf/naacl/NguyenVNNP19}
\bibinfo{author}{Nguyen\xfnm[ D.Q.]}, \bibinfo{author}{Vu\xfnm[ T.]},
  \bibinfo{author}{Nguyen\xfnm[ T.D.]}, \bibinfo{author}{Nguyen\xfnm[ D.Q.]},
  \bibinfo{author}{Phung\xfnm[ D.Q.]}.
\newblock \bibinfo{title}{A capsule network-based embedding model for knowledge
  graph completion and search personalization}.
\newblock In: \bibinfo{editor}{Burstein\xfnm[ J.]},
  \bibinfo{editor}{Doran\xfnm[ C.]}, \bibinfo{editor}{Solorio\xfnm[ T.]},
  editors. \bibinfo{booktitle}{Proceedings of the 2019 Conference of the North
  American Chapter of the Association for Computational Linguistics: Human
  Language Technologies, {NAACL-HLT} 2019, Minneapolis, MN, USA, June 2-7,
  2019, Volume 1 (Long and Short Papers)}. \bibinfo{publisher}{Association for
  Computational Linguistics}; \bibinfo{year}{2019}. p.
  \bibinfo{pages}{2180--2189}.
\newblock \URLprefix \url{https://doi.org/10.18653/v1/n19-1226}.
  \DOIprefix\doi{10.18653/v1/n19-1226}.
\bibitem[{Povey et~al.(2011)Povey, Ghoshal, Boulianne, Burget, Glembek, Goel,
  Hannemann, Motlicek, Qian, Schwarz, Silovsky, Stemmer and
  Vesely}]{Povey_ASRU2011}
\bibinfo{author}{Povey\xfnm[ D.]}, \bibinfo{author}{Ghoshal\xfnm[ A.]},
  \bibinfo{author}{Boulianne\xfnm[ G.]}, \bibinfo{author}{Burget\xfnm[ L.]},
  \bibinfo{author}{Glembek\xfnm[ O.]}, \bibinfo{author}{Goel\xfnm[ N.]},
  \bibinfo{author}{Hannemann\xfnm[ M.]}, \bibinfo{author}{Motlicek\xfnm[ P.]},
  \bibinfo{author}{Qian\xfnm[ Y.]}, \bibinfo{author}{Schwarz\xfnm[ P.]},
  \bibinfo{author}{Silovsky\xfnm[ J.]}, \bibinfo{author}{Stemmer\xfnm[ G.]},
  \bibinfo{author}{Vesely\xfnm[ K.]}.
\newblock \bibinfo{title}{The kaldi speech recognition toolkit}.
\newblock In: \bibinfo{booktitle}{IEEE 2011 Workshop on Automatic Speech
  Recognition and Understanding}. \bibinfo{publisher}{IEEE Signal Processing
  Society}; \bibinfo{year}{2011}. \bibinfo{note}{IEEE Catalog No.:
  CFP11SRW-USB}.
\bibitem[{Ramasinghe et~al.(2018)Ramasinghe, Athuraliya and
  Khan}]{DBLP:conf/eccv/RamasingheAK18}
\bibinfo{author}{Ramasinghe\xfnm[ S.]}, \bibinfo{author}{Athuraliya\xfnm[
  C.D.]}, \bibinfo{author}{Khan\xfnm[ S.H.]}.
\newblock \bibinfo{title}{A context-aware capsule network for multi-label
  classification}.
\newblock In: \bibinfo{editor}{Leal{-}Taix{\'{e}}\xfnm[ L.]},
  \bibinfo{editor}{Roth\xfnm[ S.]}, editors. \bibinfo{booktitle}{Computer
  Vision - {ECCV} 2018 Workshops - Munich, Germany, September 8-14, 2018,
  Proceedings, Part {III}}. \bibinfo{publisher}{Springer}; volume
  \bibinfo{volume}{11131} of \textit{\bibinfo{series}{Lecture Notes in Computer
  Science}}; \bibinfo{year}{2018}. p. \bibinfo{pages}{546--554}.
\newblock \URLprefix \url{https://doi.org/10.1007/978-3-030-11015-4\_40}.
  \DOIprefix\doi{10.1007/978-3-030-11015-4\_40}.
\bibitem[{Ren and Lu(2018)}]{DBLP:journals/corr/abs-1810-09177}
\bibinfo{author}{Ren\xfnm[ H.]}, \bibinfo{author}{Lu\xfnm[ H.]}.
\newblock \bibinfo{title}{Compositional coding capsule network with k-means
  routing for text classification}.
\newblock \bibinfo{journal}{CoRR}
  \bibinfo{year}{2018};\bibinfo{volume}{abs/1810.09177}.
\newblock \URLprefix \url{http://arxiv.org/abs/1810.09177}.
  \href{http://arxiv.org/abs/1810.09177}{\tt arXiv:1810.09177}.
\bibitem[{Sabour et~al.(2017)Sabour, Frosst and
  Hinton}]{DBLP:conf/nips/SabourFH17}
\bibinfo{author}{Sabour\xfnm[ S.]}, \bibinfo{author}{Frosst\xfnm[ N.]},
  \bibinfo{author}{Hinton\xfnm[ G.E.]}.
\newblock \bibinfo{title}{Dynamic routing between capsules}.
\newblock In: \bibinfo{editor}{Guyon\xfnm[ I.]}, \bibinfo{editor}{von
  Luxburg\xfnm[ U.]}, \bibinfo{editor}{Bengio\xfnm[ S.]},
  \bibinfo{editor}{Wallach\xfnm[ H.M.]}, \bibinfo{editor}{Fergus\xfnm[ R.]},
  \bibinfo{editor}{Vishwanathan\xfnm[ S.V.N.]}, \bibinfo{editor}{Garnett\xfnm[
  R.]}, editors. \bibinfo{booktitle}{Advances in Neural Information Processing
  Systems 30: Annual Conference on Neural Information Processing Systems 2017,
  4-9 December 2017, Long Beach, CA, {USA}}. \bibinfo{year}{2017}. p.
  \bibinfo{pages}{3856--3866}.
\newblock \URLprefix
  \url{http://papers.nips.cc/paper/6975-dynamic-routing-between-capsules}.
\bibitem[{Srivastava et~al.(2014)Srivastava, Hinton, Krizhevsky, Sutskever and
  Salakhutdinov}]{DBLP:journals/jmlr/SrivastavaHKSS14}
\bibinfo{author}{Srivastava\xfnm[ N.]}, \bibinfo{author}{Hinton\xfnm[ G.E.]},
  \bibinfo{author}{Krizhevsky\xfnm[ A.]}, \bibinfo{author}{Sutskever\xfnm[
  I.]}, \bibinfo{author}{Salakhutdinov\xfnm[ R.]}.
\newblock \bibinfo{title}{Dropout: a simple way to prevent neural networks from
  overfitting}.
\newblock \bibinfo{journal}{J Mach Learn Res}
  \bibinfo{year}{2014};\bibinfo{volume}{15}(\bibinfo{number}{1}):\bibinfo{pages}{1929--1958}.
\newblock \URLprefix \url{http://dl.acm.org/citation.cfm?id=2670313}.
\bibitem[{{Srivastava} et~al.(2019){Srivastava}, {Agarwal}, {Shroff} and
  {Vig}}]{8852016}
\bibinfo{author}{{Srivastava}\xfnm[ S.]}, \bibinfo{author}{{Agarwal}\xfnm[
  P.]}, \bibinfo{author}{{Shroff}\xfnm[ G.]}, \bibinfo{author}{{Vig}\xfnm[
  L.]}.
\newblock \bibinfo{title}{Hierarchical capsule based neural network
  architecture for sequence labeling}.
\newblock In: \bibinfo{booktitle}{2019 International Joint Conference on Neural
  Networks (IJCNN)}. \bibinfo{year}{2019}. p. \bibinfo{pages}{1--8}.
\bibitem[{Sun et~al.(2020)Sun, Yang, Wang, Zhang, Lin and
  Wang}]{DBLP:journals/jbi/SunYWZLW20}
\bibinfo{author}{Sun\xfnm[ C.]}, \bibinfo{author}{Yang\xfnm[ Z.]},
  \bibinfo{author}{Wang\xfnm[ L.]}, \bibinfo{author}{Zhang\xfnm[ Y.]},
  \bibinfo{author}{Lin\xfnm[ H.]}, \bibinfo{author}{Wang\xfnm[ J.]}.
\newblock \bibinfo{title}{Attention guided capsule networks for
  chemical-protein interaction extraction}.
\newblock \bibinfo{journal}{J Biomed Informatics}
  \bibinfo{year}{2020};\bibinfo{volume}{103}:\bibinfo{pages}{103392}.
\newblock \URLprefix \url{https://doi.org/10.1016/j.jbi.2020.103392}.
  \DOIprefix\doi{10.1016/j.jbi.2020.103392}.
\bibitem[{Sutskever et~al.(2014)Sutskever, Vinyals and
  Le}]{DBLP:conf/nips/SutskeverVL14}
\bibinfo{author}{Sutskever\xfnm[ I.]}, \bibinfo{author}{Vinyals\xfnm[ O.]},
  \bibinfo{author}{Le\xfnm[ Q.V.]}.
\newblock \bibinfo{title}{Sequence to sequence learning with neural networks}.
\newblock In: \bibinfo{editor}{Ghahramani\xfnm[ Z.]},
  \bibinfo{editor}{Welling\xfnm[ M.]}, \bibinfo{editor}{Cortes\xfnm[ C.]},
  \bibinfo{editor}{Lawrence\xfnm[ N.D.]}, \bibinfo{editor}{Weinberger\xfnm[
  K.Q.]}, editors. \bibinfo{booktitle}{Advances in Neural Information
  Processing Systems 27: Annual Conference on Neural Information Processing
  Systems 2014, December 8-13 2014, Montreal, Quebec, Canada}.
  \bibinfo{year}{2014}. p. \bibinfo{pages}{3104--3112}.
\newblock \URLprefix
  \url{http://papers.nips.cc/paper/5346-sequence-to-sequence-learning-with-neural-networks}.
\bibitem[{Tsai et~al.(2020)Tsai, Srivastava, Goh and
  Salakhutdinov}]{DBLP:conf/iclr/TsaiSGS20}
\bibinfo{author}{Tsai\xfnm[ Y.H.]}, \bibinfo{author}{Srivastava\xfnm[ N.]},
  \bibinfo{author}{Goh\xfnm[ H.]}, \bibinfo{author}{Salakhutdinov\xfnm[ R.]}.
\newblock \bibinfo{title}{Capsules with inverted dot-product attention
  routing}.
\newblock In: \bibinfo{booktitle}{8th International Conference on Learning
  Representations, {ICLR} 2020, Addis Ababa, Ethiopia, April 26-30, 2020}.
  \bibinfo{publisher}{OpenReview.net}; \bibinfo{year}{2020}. \URLprefix
  \url{https://openreview.net/forum?id=HJe6uANtwH}.
\bibitem[{Vaswani et~al.(2017)Vaswani, Shazeer, Parmar, Uszkoreit, Jones,
  Gomez, Kaiser and Polosukhin}]{DBLP:conf/nips/VaswaniSPUJGKP17}
\bibinfo{author}{Vaswani\xfnm[ A.]}, \bibinfo{author}{Shazeer\xfnm[ N.]},
  \bibinfo{author}{Parmar\xfnm[ N.]}, \bibinfo{author}{Uszkoreit\xfnm[ J.]},
  \bibinfo{author}{Jones\xfnm[ L.]}, \bibinfo{author}{Gomez\xfnm[ A.N.]},
  \bibinfo{author}{Kaiser\xfnm[ L.]}, \bibinfo{author}{Polosukhin\xfnm[ I.]}.
\newblock \bibinfo{title}{Attention is all you need}.
\newblock In: \bibinfo{editor}{Guyon\xfnm[ I.]}, \bibinfo{editor}{von
  Luxburg\xfnm[ U.]}, \bibinfo{editor}{Bengio\xfnm[ S.]},
  \bibinfo{editor}{Wallach\xfnm[ H.M.]}, \bibinfo{editor}{Fergus\xfnm[ R.]},
  \bibinfo{editor}{Vishwanathan\xfnm[ S.V.N.]}, \bibinfo{editor}{Garnett\xfnm[
  R.]}, editors. \bibinfo{booktitle}{Advances in Neural Information Processing
  Systems 30: Annual Conference on Neural Information Processing Systems 2017,
  4-9 December 2017, Long Beach, CA, {USA}}. \bibinfo{year}{2017}. p.
  \bibinfo{pages}{5998--6008}.
\newblock \URLprefix
  \url{http://papers.nips.cc/paper/7181-attention-is-all-you-need}.
\bibitem[{Vesperini et~al.(2019)Vesperini, Gabrielli, Principi and
  Squartini}]{DBLP:journals/jstsp/VesperiniGPS19}
\bibinfo{author}{Vesperini\xfnm[ F.]}, \bibinfo{author}{Gabrielli\xfnm[ L.]},
  \bibinfo{author}{Principi\xfnm[ E.]}, \bibinfo{author}{Squartini\xfnm[ S.]}.
\newblock \bibinfo{title}{Polyphonic sound event detection by using capsule
  neural networks}.
\newblock \bibinfo{journal}{J Sel Topics Signal Processing}
  \bibinfo{year}{2019};\bibinfo{volume}{13}(\bibinfo{number}{2}):\bibinfo{pages}{310--322}.
\newblock \URLprefix \url{https://doi.org/10.1109/JSTSP.2019.2902305}.
  \DOIprefix\doi{10.1109/JSTSP.2019.2902305}.
\bibitem[{Wang and Liu(2018)}]{DBLP:conf/iclr/Wang018}
\bibinfo{author}{Wang\xfnm[ D.]}, \bibinfo{author}{Liu\xfnm[ Q.]}.
\newblock \bibinfo{title}{An optimization view on dynamic routing between
  capsules}.
\newblock In: \bibinfo{booktitle}{6th International Conference on Learning
  Representations, {ICLR} 2018, Vancouver, BC, Canada, April 30 - May 3, 2018,
  Workshop Track Proceedings}. \bibinfo{publisher}{OpenReview.net};
  \bibinfo{year}{2018}. \URLprefix
  \url{https://openreview.net/forum?id=HJjtFYJDf}.
\bibitem[{Wang(2019)}]{wang-2019-towards}
\bibinfo{author}{Wang\xfnm[ M.]}.
\newblock \bibinfo{title}{Towards linear time neural machine translation with
  capsule networks}.
\newblock In: \bibinfo{booktitle}{Proceedings of the 2019 Conference on
  Empirical Methods in Natural Language Processing and the 9th International
  Joint Conference on Natural Language Processing (EMNLP-IJCNLP)}.
  \bibinfo{address}{Hong Kong, China}: \bibinfo{publisher}{Association for
  Computational Linguistics}; \bibinfo{year}{2019}. p.
  \bibinfo{pages}{803--812}.
\newblock \URLprefix \url{https://www.aclweb.org/anthology/D19-1074}.
  \DOIprefix\doi{10.18653/v1/D19-1074}.
\bibitem[{Wu et~al.(2019)Wu, Liu, Cao, Li, Yu, Dai, Ma, Hu, Wu, Liu and
  Meng}]{DBLP:conf/icassp/WuLCLYDMHWLM19}
\bibinfo{author}{Wu\xfnm[ X.]}, \bibinfo{author}{Liu\xfnm[ S.]},
  \bibinfo{author}{Cao\xfnm[ Y.]}, \bibinfo{author}{Li\xfnm[ X.]},
  \bibinfo{author}{Yu\xfnm[ J.]}, \bibinfo{author}{Dai\xfnm[ D.]},
  \bibinfo{author}{Ma\xfnm[ X.]}, \bibinfo{author}{Hu\xfnm[ S.]},
  \bibinfo{author}{Wu\xfnm[ Z.]}, \bibinfo{author}{Liu\xfnm[ X.]},
  \bibinfo{author}{Meng\xfnm[ H.]}.
\newblock \bibinfo{title}{Speech emotion recognition using capsule networks}.
\newblock In: \bibinfo{booktitle}{{IEEE} International Conference on Acoustics,
  Speech and Signal Processing, {ICASSP} 2019, Brighton, United Kingdom, May
  12-17, 2019}. \bibinfo{publisher}{{IEEE}}; \bibinfo{year}{2019}. p.
  \bibinfo{pages}{6695--6699}.
\newblock \URLprefix \url{https://doi.org/10.1109/ICASSP.2019.8683163}.
  \DOIprefix\doi{10.1109/ICASSP.2019.8683163}.
\bibitem[{Xia et~al.(2018)Xia, Zhang, Yan, Chang and
  Yu}]{DBLP:conf/emnlp/XiaZYCY18}
\bibinfo{author}{Xia\xfnm[ C.]}, \bibinfo{author}{Zhang\xfnm[ C.]},
  \bibinfo{author}{Yan\xfnm[ X.]}, \bibinfo{author}{Chang\xfnm[ Y.]},
  \bibinfo{author}{Yu\xfnm[ P.S.]}.
\newblock \bibinfo{title}{Zero-shot user intent detection via capsule neural
  networks}.
\newblock In: \bibinfo{editor}{Riloff\xfnm[ E.]}, \bibinfo{editor}{Chiang\xfnm[
  D.]}, \bibinfo{editor}{Hockenmaier\xfnm[ J.]}, \bibinfo{editor}{Tsujii\xfnm[
  J.]}, editors. \bibinfo{booktitle}{Proceedings of the 2018 Conference on
  Empirical Methods in Natural Language Processing, Brussels, Belgium, October
  31 - November 4, 2018}. \bibinfo{publisher}{Association for Computational
  Linguistics}; \bibinfo{year}{2018}. p. \bibinfo{pages}{3090--3099}.
\newblock \URLprefix \url{https://doi.org/10.18653/v1/d18-1348}.
  \DOIprefix\doi{10.18653/v1/d18-1348}.
\bibitem[{Xinyi and Chen(2019)}]{DBLP:conf/iclr/XinyiC19}
\bibinfo{author}{Xinyi\xfnm[ Z.]}, \bibinfo{author}{Chen\xfnm[ L.]}.
\newblock \bibinfo{title}{Capsule graph neural network}.
\newblock In: \bibinfo{booktitle}{7th International Conference on Learning
  Representations, {ICLR} 2019, New Orleans, LA, USA, May 6-9, 2019}.
  \bibinfo{publisher}{OpenReview.net}; \bibinfo{year}{2019}. \URLprefix
  \url{https://openreview.net/forum?id=Byl8BnRcYm}.
\bibitem[{Yan(2018)}]{xiongyan2018master}
\bibinfo{author}{Yan\xfnm[ X.]}.
\newblock \bibinfo{title}{Using Capsule Networks for Image and Speech
  Recognition Problems}.
\newblock Master's thesis; Arizona State University, Electrical engineering;
  \bibinfo{year}{2018}.
\bibitem[{Zhang et~al.(2019{\natexlab{a}})Zhang, Li, Du, Fan and
  Yu}]{DBLP:conf/acl/ZhangLDFY19}
\bibinfo{author}{Zhang\xfnm[ C.]}, \bibinfo{author}{Li\xfnm[ Y.]},
  \bibinfo{author}{Du\xfnm[ N.]}, \bibinfo{author}{Fan\xfnm[ W.]},
  \bibinfo{author}{Yu\xfnm[ P.S.]}.
\newblock \bibinfo{title}{Joint slot filling and intent detection via capsule
  neural networks}.
\newblock In: \bibinfo{editor}{Korhonen\xfnm[ A.]},
  \bibinfo{editor}{Traum\xfnm[ D.R.]}, \bibinfo{editor}{M{\`{a}}rquez\xfnm[
  L.]}, editors. \bibinfo{booktitle}{Proceedings of the 57th Conference of the
  Association for Computational Linguistics, {ACL} 2019, Florence, Italy, July
  28- August 2, 2019, Volume 1: Long Papers}. \bibinfo{publisher}{Association
  for Computational Linguistics}; \bibinfo{year}{2019}{\natexlab{a}}. p.
  \bibinfo{pages}{5259--5267}.
\newblock \URLprefix \url{https://doi.org/10.18653/v1/p19-1519}.
  \DOIprefix\doi{10.18653/v1/p19-1519}.
\bibitem[{Zhang et~al.(2018)Zhang, Deng, Sun, Chen, Zhang and
  Chen}]{zhang-etal-2018-attention}
\bibinfo{author}{Zhang\xfnm[ N.]}, \bibinfo{author}{Deng\xfnm[ S.]},
  \bibinfo{author}{Sun\xfnm[ Z.]}, \bibinfo{author}{Chen\xfnm[ X.]},
  \bibinfo{author}{Zhang\xfnm[ W.]}, \bibinfo{author}{Chen\xfnm[ H.]}.
\newblock \bibinfo{title}{Attention-based capsule networks with dynamic routing
  for relation extraction}.
\newblock In: \bibinfo{booktitle}{Proceedings of the 2018 Conference on
  Empirical Methods in Natural Language Processing}.
  \bibinfo{address}{Brussels, Belgium}: \bibinfo{publisher}{Association for
  Computational Linguistics}; \bibinfo{year}{2018}. p.
  \bibinfo{pages}{986--992}.
\newblock \URLprefix \url{https://www.aclweb.org/anthology/D18-1120}.
  \DOIprefix\doi{10.18653/v1/D18-1120}.
\bibitem[{Zhang et~al.(2019{\natexlab{b}})Zhang, Zhou and
  Wu}]{DBLP:series/sci/Zhang0W19}
\bibinfo{author}{Zhang\xfnm[ S.]}, \bibinfo{author}{Zhou\xfnm[ Q.]},
  \bibinfo{author}{Wu\xfnm[ X.]}.
\newblock \bibinfo{title}{Fast dynamic routing based on weighted kernel density
  estimation}.
\newblock In: \bibinfo{editor}{Lu\xfnm[ H.]}, editor.
  \bibinfo{booktitle}{Cognitive Internet of Things: Frameworks, Tools and
  Applications}. \bibinfo{publisher}{Springer}; volume \bibinfo{volume}{810} of
  \textit{\bibinfo{series}{Studies in Computational Intelligence}};
  \bibinfo{year}{2019}{\natexlab{b}}. p. \bibinfo{pages}{301--309}.
\newblock \URLprefix \url{https://doi.org/10.1007/978-3-030-04946-1\_30}.
  \DOIprefix\doi{10.1007/978-3-030-04946-1\_30}.
\bibitem[{Zhang et~al.(2016)Zhang, Pezeshki, Brakel, Zhang, Laurent, Bengio and
  Courville}]{DBLP:conf/interspeech/ZhangPBZLBC16}
\bibinfo{author}{Zhang\xfnm[ Y.]}, \bibinfo{author}{Pezeshki\xfnm[ M.]},
  \bibinfo{author}{Brakel\xfnm[ P.]}, \bibinfo{author}{Zhang\xfnm[ S.]},
  \bibinfo{author}{Laurent\xfnm[ C.]}, \bibinfo{author}{Bengio\xfnm[ Y.]},
  \bibinfo{author}{Courville\xfnm[ A.C.]}.
\newblock \bibinfo{title}{Towards end-to-end speech recognition with deep
  convolutional neural networks}.
\newblock In: \bibinfo{editor}{Morgan\xfnm[ N.]}, editor.
  \bibinfo{booktitle}{Interspeech 2016, 17th Annual Conference of the
  International Speech Communication Association, San Francisco, CA, USA,
  September 8-12, 2016}. \bibinfo{publisher}{{ISCA}}; \bibinfo{year}{2016}. p.
  \bibinfo{pages}{410--414}.
\newblock \URLprefix \url{https://doi.org/10.21437/Interspeech.2016-1446}.
  \DOIprefix\doi{10.21437/Interspeech.2016-1446}.
\bibitem[{Zhao et~al.(2019)Zhao, Kleinhans, Sandhu, Patel and
  Unnikrishnan}]{DBLP:journals/corr/abs-1903-09662}
\bibinfo{author}{Zhao\xfnm[ Z.]}, \bibinfo{author}{Kleinhans\xfnm[ A.]},
  \bibinfo{author}{Sandhu\xfnm[ G.]}, \bibinfo{author}{Patel\xfnm[ I.]},
  \bibinfo{author}{Unnikrishnan\xfnm[ K.P.]}.
\newblock \bibinfo{title}{Capsule networks with max-min normalization}.
\newblock \bibinfo{journal}{CoRR}
  \bibinfo{year}{2019};\bibinfo{volume}{abs/1903.09662}.
\newblock \URLprefix \url{http://arxiv.org/abs/1903.09662}.
  \href{http://arxiv.org/abs/1903.09662}{\tt arXiv:1903.09662}.
\bibitem[{Zhong et~al.(2020)Zhong, Liu, Li, Chen, Lu, Dong, Wu and
  Zhong}]{DBLP:journals/nca/ZhongLLCLDWZ20}
\bibinfo{author}{Zhong\xfnm[ X.]}, \bibinfo{author}{Liu\xfnm[ J.]},
  \bibinfo{author}{Li\xfnm[ L.]}, \bibinfo{author}{Chen\xfnm[ S.]},
  \bibinfo{author}{Lu\xfnm[ W.]}, \bibinfo{author}{Dong\xfnm[ Y.]},
  \bibinfo{author}{Wu\xfnm[ B.]}, \bibinfo{author}{Zhong\xfnm[ L.]}.
\newblock \bibinfo{title}{An emotion classification algorithm based on
  spt-capsnet}.
\newblock \bibinfo{journal}{Neural Computing and Applications}
  \bibinfo{year}{2020};\bibinfo{volume}{32}(\bibinfo{number}{7}):\bibinfo{pages}{1823--1837}.
\newblock \URLprefix \url{https://doi.org/10.1007/s00521-019-04621-y}.
  \DOIprefix\doi{10.1007/s00521-019-04621-y}.

\end{thebibliography}

\end{document}